\documentclass[reprint,amsmath,amssymb,aps]{revtex4-2}
\usepackage{xcolor}
\usepackage{graphicx}
\usepackage{dcolumn}
\usepackage{bm}
\usepackage{physics}
\usepackage{mathtools}
\usepackage{enumerate}
\newcommand{\om}{\omega}
\newcommand{\Om}{\Omega}
\newcommand{\bpsi}{\Bar{\psi}}
\newcommand{\bx}{\mathbf{x}}
\newcommand{\by}{\mathbf{y}}
\newcommand{\bk}{\mathbf{k}}
\newcommand{\bp}{\mathbf{p}}
\newcommand{\bq}{\mathbf{q}}
\newcommand{\up}{\uparrow}
\newcommand{\down}{\downarrow}
\newcommand{\gap}{\Delta}

\newcommand{\bark}{\Bar{k}}
\newcommand{\barq}{\Bar{q}}
\newcommand{\bom}{\Bar{\omega}}
\newcommand{\bmu}{\Bar{\mu}}
\newcommand{\bxi}{\Bar{\xi}}
\newcommand{\bgap}{\Bar{\Delta}}
\newcommand{\bE}{\bar{E}}

\begin{document}

\preprint{APS/123-QED}

\title{Following the Higgs mode across the BCS-BEC crossover in two dimensions}

\author{Dan Phan}
\author{Andrey V. Chubukov}
\affiliation{
School of Physics and Astronomy, University of Minnesota, Minneapolis, Minnesota 55455, USA
}

\date{\today}
\begin{abstract}
Although substantial effort has been dedicated to analyzing the Higgs (amplitude) mode in superconducting systems, there are relatively few studies of the Higgs peak's dispersion and width, quantities which are observable in spectroscopic measurements. These properties can be obtained from the location of the pole of the longitudinal (Higgs) susceptibility in the lower half-plane of complex frequency. We analyze the behavior of the Higgs mode in a 2D neutral fermionic superfluid at $T=0$ throughout the crossover from Bardeen-Cooper-Schrieffer (BCS) to Bose-Einstein condensation (BEC) behavior. This occurs when the dressed chemical potential $\mu$ changes sign from positive to negative. For $\mu >0$, we find a pole in the Higgs susceptibility in the lower half-plane of frequency and demonstrate that it leads to a well-defined peak in spectroscopic probes. For $\mu<0$, the pole still exists, but is ``hidden,'' not giving rise to a peak in spectroscopic probes. Extending this analysis to a charged superconductor, we find that the Higgs mode is unaffected by the long-range Coulomb interaction.
\end{abstract}
\maketitle

 \section*{\uppercase{Introduction}}
\label{sec:intro}

Superconducting and superfluid phases of interacting fermions are characterized by spontaneous breaking of U(1) gauge symmetry, resulting in a nonzero complex order parameter $\gap= \abs{\gap}e^{i\varphi}$. Fluctuations in this order parameter can be decomposed into fluctuations of the phase $\varphi$ (the Anderson-Bogoliubov-Goldstone or ABG mode), and the amplitude $\abs{\gap}$ (the Higgs mode) \cite{Anderson1958,Anderson1963,Littlewood1982,Ohashi1998}. The Higgs mode has traditionally been difficult to observe experimentally, since as a scalar field, it does not couple linearly to
the electromagnetic field \cite{Shimano2020}. Indeed, until recently, the only clear experimental observation of the Higgs mode has been in 2$H$-NbSe$_2$ \cite{Sooryakumar1980,Sooryakumar1981}, due to the coexistence of charge-density wave order and superconductivity \cite{Littlewood1981,Littlewood1982,Benfatto_4}. However, in the past decade, advances in ultra-fast THz and Raman spectroscopy have led to numerous reports of observations of the Higgs mode \cite{Pekker2015,Assa,Assa_1,Matsunaga2014,Matsunaga2017,Katsumi2018,Yang2019,Chu2020,Katsumi2020,Grasset2022, Benfatto_1,Maiti,Cea2015,Puviani,*Benfatto_2,*Puviani_1,Manske,Gallais,Gallais_1,*Gallais_2,Sacuto,Grasset2018}.

In a 3D $s$-wave superconductor where the Fermi energy is much larger than the gap ($E_F \gg \gap$), the Higgs mode has frequency $\om_H = 2\gap$ in the long-wavelength limit, $\bq=0$. As such, the Higgs mode lies on the edge of the two-particle continuum, where Cooper pairs break up into two Bogoliubov quasiparticles \cite{varma2002higgs}. One consequence of this can be seen in the longitudinal (Higgs) susceptibility describing amplitude oscillations,
which exhibits a branch cut rather than a pole,
$\chi_H(\om+i\delta,\bq=0) \sim 1/\sqrt{\om^2-4\gap^2}$
~\cite{Cea2015}. This square-root singularity leads to amplitude oscillations which decay in time as a power law \cite{Volkov1974}, as opposed to the exponential decay one expects from a true pole.

\begin{figure}
    \centering
    \includegraphics[width=\columnwidth]{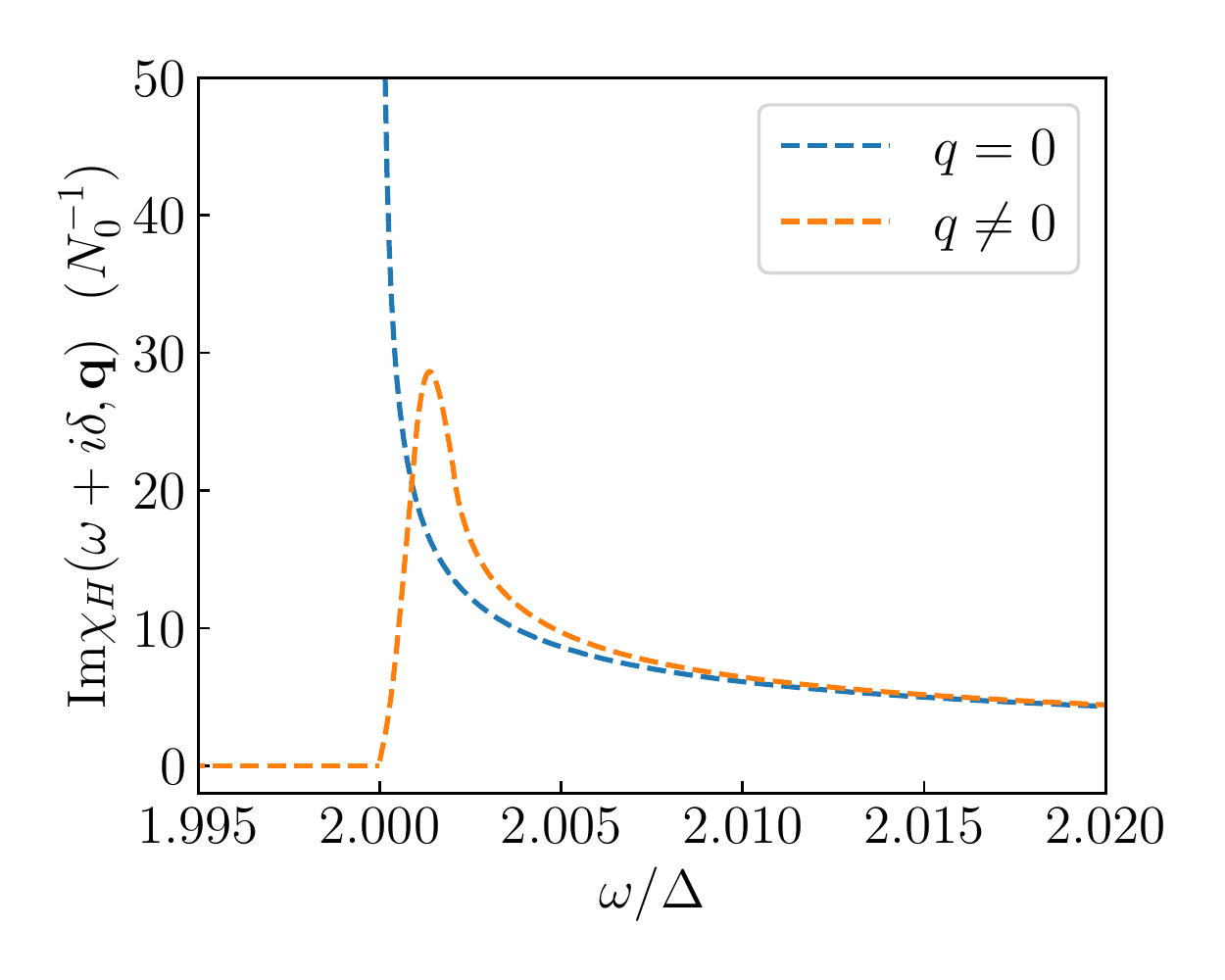}
    \caption{Behavior of the Higgs susceptibility in the BCS regime, as a function of frequency at zero and nonzero $q$. The two-particle continuum begins at $\om=2\gap$.}
    \label{fig:chi_peak}
\end{figure}

\begin{figure}
    \centering
    \includegraphics[width=\columnwidth]{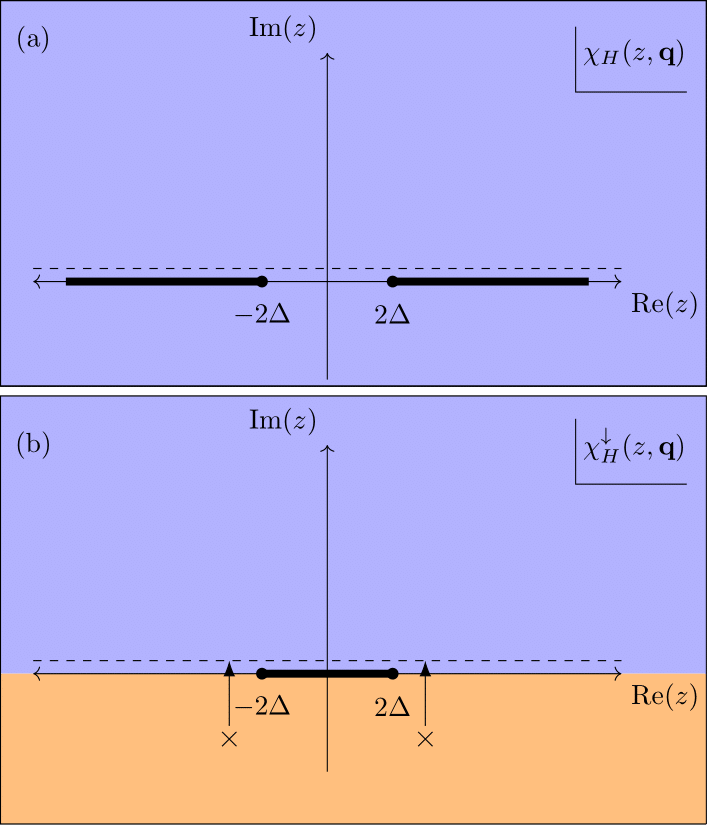}
    \caption{The branch cut structure of (a) the Higgs susceptibility $\chi_H(z,\bq)$ and (b) its analytical continuation $\chi_H^\down(z,\bq)$ in the complex $z$ plane in the BCS regime, $\mu >0$. In both panels, the dashed line
    indicates $z = \om+i\delta$, the frequencies which are probed in spectroscopic experiments. The background coloring denotes Riemann sheets, on which $\chi_H(z,\bq)$ and $\chi_H^\down(z,\bq)$ are defined. The function $\chi_H(z,\bq)$ is defined on one Riemann sheet throughout the complex plane, while $\chi_H^\down(z,\bq)$ is defined on two Riemann sheets, depending on the sign of $\Im(z)$.
    Crosses in panel (b) show the position of poles in $\chi_H^\down(z,\bq)$. Due to the analyticity of $\chi_H^\down(z,\bq)$ across the real axis for $\om > 2\Delta$, these poles lead to an experimentally observable peak in $\Im\chi_H(\om+i\delta,\bq)$. To highlight this, we have added vertical arrows connecting these poles to the dashed line above the real axis. There are additional branch cuts in $\chi_H^\down(z,\bq)$ at larger $\abs{\om}$, which we do not present here (see Sec. \ref{sec:ac_overview}). The presence of these additional branch cuts limits the frequency range where the poles of $\chi_H^\down(z,\bq)$ give rise to peaks in $\Im\chi_H(\om+i\delta,\bq)$. }
    \label{fig:branch_cut_pos_mu}
\end{figure}

At nonzero $\bq$, the square-root singularity disappears, and the spectral function $\Im\chi_H(\om+i\delta,\bq)$ develops a peak at a frequency $\om$ above $2\gap$, whose width is small but finite (see Fig. \ref{fig:chi_peak}). It is natural to assume that a narrow peak at a frequency immediately above the real axis can be understood as resulting from a pole in  $\chi_H(z,\bq)$ at a complex frequency $z_\bq = \omega' + i \omega^{''}$ slightly below the real axis. This is not guaranteed however, as the presence of the two-particle continuum implies that the function $\chi_H(z,\bq)$ has branch cuts along the real-frequency axis for $\abs{\om} > 2\gap$ (see Fig. \ref{fig:branch_cut_pos_mu}(a).) This branch cut implies that the behavior of $\chi_H(z,\bq)$ below the real axis (for e.g. $\om = \Re(z) > 2\gap$) is not smoothly connected to the behavior of $\chi_H(\om+i\delta,\bq)$. In contrast, $\chi_H(z,\bq)$ is smoothly connected to $\chi_H(\om+i\delta,\bq)$ for $\om < 2\gap$, as there is no branch cut along the real axis in this case.

This line of reasoning suggests that more careful analysis is needed to determine whether the presence of a peak in $\Im\chi_H(\om+i\delta,\bq)$ at $\om>2\gap$ arises from a pole in $\chi_H(z,\bq)$ in the lower half-plane, and conversely, whether the absence of such a peak implies that there is no pole in $\chi_H(z,\bq)$. To address this issue, one has to \textit{analytically continue} $\chi_H(z,\bq)$ through the branch cut into the lower half-plane, and check whether this analytically continued susceptibility has a pole. We denote this function $\chi_H^\down(z,\bq)$ below. It is equal to $\chi_H(z,\bq)$ in the upper half-plane and is constructed to be smooth across the real-frequency axis for $\abs{\om} > 2\gap$. The branch cut structure of $\chi_H(z,\bq)$ and its analytical continuation $\chi_H^\down(z,\bq)$ are illustrated in Fig. \ref{fig:branch_cut_pos_mu}(a) and Fig. \ref{fig:branch_cut_pos_mu}(b), respectively.

Since $\chi_H^\down(z,\bq)$ is analytic across the real axis for $\abs{\om}>2\gap$, a pole in $\chi_H^\down(z,\bq)$ at a frequency $z_\bq$ close to the real axis, with $\Re(z_\bq) > 2\gap$, necessarily leads to a peak in $\Im\chi_H^\down(z,\bq)$ immediately above the real axis. To highlight
this, we have added in Fig. \ref{fig:branch_cut_pos_mu}(b) vertical arrows from the positions of the poles (shown as crosses in the lower half-plane) to $z = \om+i\delta$. Henceforth, we refer to such poles as Higgs modes.

This reasoning does not hold for poles with $\Re(z_\bq) < 2\gap$, due to the \textit{non-analyticity} of $\chi_H^\down(z,\bq)$ across the real axis for such $\Re(z_\bq)$. In this case, there is no peak in $\Im\chi_H(\om+i\delta,\bq)$. The situation is similar to that of zero-sound collective modes in 2D for small, negative values of the Landau parameter $F_0$: the charge susceptibility has a pole in the lower half-plane of frequency, but does not give rise to a peak in the spectral function due to a branch cut across the real axis~\cite{Klein2020}. Borrowing the notation from that paper, we refer to such a pole as a hidden mode.

From the perspective of complex analysis, it is natural to think of $\chi_H(z,\bq)$ and $\chi_H^\down(z,\bq)$ as components of a single function defined on a Riemann surface, which consists of multiple Riemann sheets glued together along the real axis. From this perspective, the discontinuity in $\chi_H(z,\bq)$ across the real axis for $\abs{\om} > 2\gap$ is a consequence of staying on the same Riemann sheet as we cross the real axis. Similarly, the smooth evolution of $\chi_H^\down(\om,\bq)$ across the real axis is obtained by transitioning from one Riemann sheet at $\Im(z) >0$ to another at $\Im(z)<0$
\footnote{In fact, there are complications to this procedure. There is no way to construct a function which is analytic across the real axis for all $\abs{\om}>2\gap$. Here, one should think of $\chi_H^\down(z,\bq)$ as being analytic for $\abs{\om} \in (2\gap,\om_2)$ for some frequency $\om_2$. We discuss the analytic continuation in more detail in Sec. \ref{sec:ac_overview}.}.
We illustrate this in Fig. \ref{fig:branch_cut_pos_mu} via the background coloring: different coloring in Fig. \ref{fig:branch_cut_pos_mu}(b) indicates that $\chi^\down_H(z,\bq)$ lives on different Riemann sheets in the upper and lower half-planes.

In this respect, the pole in $\chi^\down_H(z,\bq)$ exists on an unphysical Riemann sheet, different from the physical Riemann sheet where $\Im\chi^\down_H(\omega+i \delta,\bq)$ is measured in spectroscopic probes \cite{Behrle2018,Sobirey2022}. However, due to the analyticity of $\chi_H^\down(z,\bq)$ across the real axis for $\om > 2\gap$, poles on this unphysical Riemann sheet lead to observable peaks in the spectral function $\Im\chi_H^\down(\om+i\delta,\bq)$. Such poles have been referred to as mirage modes in Ref. \cite{Klein2020}.

We re-iterate that a mirage mode with $\abs{\om}>2\gap$ on the unphysical Riemann sheet, if it exists, gives rise to a measurable peak in $\Im\chi_H(\omega+ i \delta,\bq)$. This is due to the analyticity of $\chi_H^\down(z,\bq)$ for $\abs{\om}> 2\gap$ along the vertical path connecting the pole at $z_\bq$ in the lower half-plane on an unphysical Riemann sheet, to the frequency $z = \Re(z_\bq)+i\delta$ in the upper half-plane on the physical Riemann sheet.

On the other hand, if a pole of $\chi_H^\down(z,\bq)$ on the unphysical Riemann sheet has $\Re(z_\bq) < 2\gap$, it is no longer smoothly connected to the spectral function $\Im\chi_H(\om+i\delta,\bq)$ on the physical Riemann sheet. Instead, the pole at $z_\bq$ leads to a peak in the spectral function evaluated on a different, unphysical Riemann sheet.
The pole with $\Re(z_\bq) < 2\gap$ then becomes a hidden mode.

In 3D, the analytic structure of $\chi_H^\down(z,\bq)$ has been analyzed by Andrianov and Popov \cite{Andrianov1976}. In the high-density BCS limit, where the chemical potential $\mu$ is much larger than the gap $\gap$, they found that a pole in $\chi_H^\down(z,\bq)$ does exist, and its location is $z_\bq= 2\gap + (0.2369 -0.2956i)\frac{q^2}{2m}\frac{\mu}{\gap}$. We note that this result for $\Re(z_\bq)$ disagrees with the commonly-cited result for the Higgs mode dispersion, which in our notation reads $z_\bq = 2\Delta + \frac{1}{3}\frac{q^2}{2m}\frac{\mu}{\gap} - i \frac{\pi^2}{12}\sqrt{\frac{\mu}{2m}}q$ \cite{Littlewood1982}. We discuss the reason for this disagreement in Sec. \ref{app:technical_details} of the Supplementary Information (SI).

Away from the high-density limit, particle-hole symmetry disappears, leading to a coupling of the amplitude and phase fluctuations. Previous work~\cite{Pekker2015,Cea2015} has suggested that this loss of particle-hole symmetry eventually leads to the disappearance of the Higgs mode. An analysis of this has been done recently by Kurkjian et al. \cite{Kurkjian2019}. Writing the dispersion of the Higgs mode as $z_\bq = 2\gap + \zeta \frac{q^2}{2m}\frac{\mu}{\gap}$, they demonstrated that $\Re(\zeta)$ decreases from its high-density value of $\Re(\zeta) = 0.2369$ as one lowers the chemical potential $\mu$, changes sign at $\mu_c \approx 0.8267\gap$, and becomes negative for smaller $\mu$. For $0< \mu < \mu_c$, when the system approaches the BEC regime, the pole has $\Re(z_\bq) < 2\gap$. The authors then argued that this pole does not give rise to a peak in $\Im \chi_H(\omega+ i \delta,\bq)$.

In this work, we extend the analysis by Kurkjian et al. to 2D, investigating the dispersion and damping rate of the Higgs mode in a neutral fermionic superfluid and a charged superconductor as a function of density at $T=0$. In contrast to the 3D case where $\mu$ and $\gap$ become comparable only at strong coupling, in 2D one can tune between the BCS and BEC regimes already at weak coupling by varying the fermionic density. This is because for a parabolic dispersion in 2D, a two-fermion bound state exists even at arbitrarily weak attraction \cite{Randeria1989}. For values of $E_F$ larger than $E_0$ (where $E_0$ is half the bound-state energy of two fermions in vacuum), the system is in the BCS regime ($\mu \gtrsim 0$). In the low-density limit where $E_F \ll E_0$, the system is instead in the BEC regime. Here, the chemical potential $\mu$ is strongly renormalized down from its normal state value $E_F$ and becomes negative, $\mu \approx -E_0$.

In the high-density BCS limit, we find a pole in $\chi_H^\down(z,\bq)$ at $z_\bq = 2\Delta + (0.5 -0.4308 i) \frac{q^2}{2m}\frac{\mu}{\gap}$. As $\mu/\Delta$ decreases, the pole moves to $z_\bq = 2\Delta + (0.5 - i \beta)\frac{q^2}{2m}\frac{\mu}{\gap}$, where $\beta$ interpolates between $0.4308$ at $\mu \gg \gap$ and the much larger $\frac{e}{16}\sqrt{\frac{2\gap}{\mu}}$ at $\mu \ll \gap$. We also calculate the residue of the pole, finding that it scales linearly with $q$, as in 3D \cite{Kurkjian2019}. We next move away from the long-wavelength limit and trace the position of the pole in $\chi_H^\down(z,\bq)$ as a function of $q$. We find that the Higgs mode quickly becomes heavily damped with increasing $q$.

\begin{figure}
    \centering
    \includegraphics[width=\columnwidth]{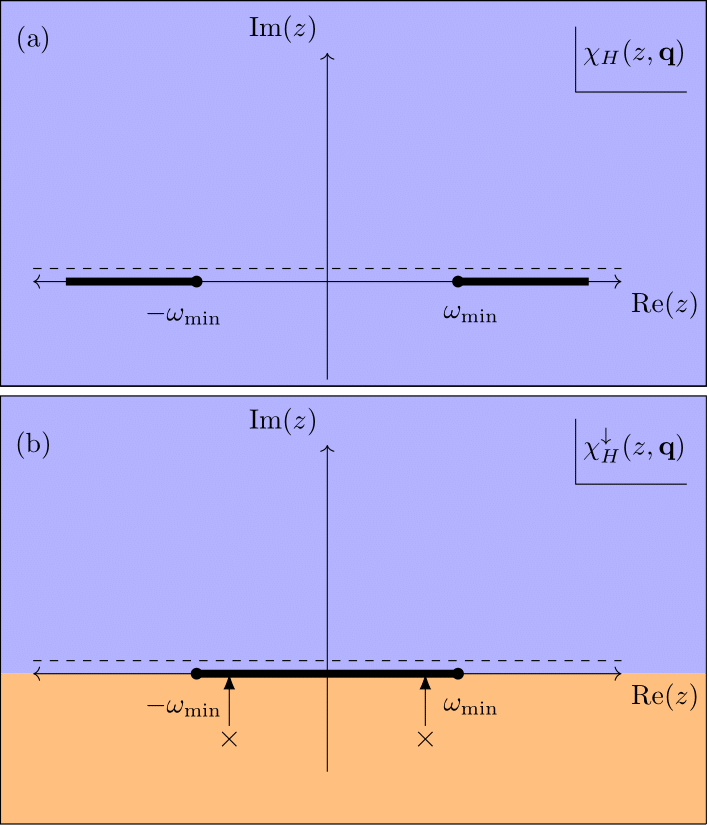}
    \caption{The branch cut structure of (a) the Higgs susceptibility $\chi_H(z,\bq)$ and (b) its analytical continuation $\chi_H^\down(z,\bq)$ in the complex $z$ plane, in the BEC regime, where $\mu < 0$. In both panels, the dashed line
    indicates frequencies which are probed in spectroscopic experiments.
    The background color denotes Riemann sheets, and highlights that $\chi_H(z,\bq)$ in panel (a) is defined on one Riemann sheet throughout the complex plane. In contrast, $\chi_H^\down(z,\bq)$ is defined on two Riemann sheets, depending on the sign of $\Im(z)$.
    The crosses in panel (b) indicate the position of
    the poles of $\chi_H^\down(z,\bq)$. Unlike when $\mu > 0$, the poles lie below the threshold of the two-particle-continuum at $\omega_\mathrm{min} = 2 \sqrt{\Delta^2 + (\abs{\mu}+q^2/8m)^2}$. The function $\chi_H^\down(\om,\bq)$ has a branch cut at $\abs{\om} < \omega_\mathrm{min}$, and the poles do not lead to observable peaks in $\Im \chi_H(\om+i\delta,\bq)$. This is emphasized with the vertical arrows-- the poles of $\chi_H^\down(z,\bq)$ are obstructed from leading to a peak above the real axis due to the discontinuity across the branch cut. }
    \label{fig:branch_cut_neg_mu}
\end{figure}

Crossing from the BCS regime ($\mu > 0$) to the BEC regime ($\mu < 0$), we find that the Higgs mode becomes hidden for $\mu < \mu_c = 0$. We illustrate the situation in the BEC regime in Fig. \ref{fig:branch_cut_neg_mu}, where the branch cut structure of $\chi_H(z,\bq)$ and $\chi_H^\down(z,\bq)$ is presented in panels (a) and (b), respectively. Here, the branch points at $\pm 2\gap$ have been replaced with $\pm \om_\mathrm{min}\equiv \pm
2\sqrt{\gap^2+(\abs{\mu}+q^2/8m)^2}$, which is the lower bound of the two-particle continuum for $\mu < 0$. As in the case when $\mu > 0$, $\chi_H^\down(z,\bq)$ has poles in the lower-half plane. However, we find that these poles are hidden below the branch cut in $\chi_H^\down(z,\bq)$, extending from $-\om_\mathrm{min}$ to $+\om_\mathrm{min}$. Hence, as in 3D, the poles do not give rise to spectroscopic signatures at frequencies immediately above the real axis.

We then investigate how the dispersion and damping rate of the Higgs mode is modified by the inclusion of the long-range Coulomb interaction. Our calculations show that the dispersion, damping rate, and residue of the Higgs mode is \textit{unchanged} from that of the neutral superfluid. We find that this is true in both two and three dimensions.

The paper is structured as follows: In Secs. \ref{sec:mean_field} and \ref{sec:gaussian_fluctuations}, we review how we obtain the Higgs susceptibility, using the functional-integral method within the Gaussian approximation. In Sec. \ref{sec:neutral_analytic}, we then obtain the small-$q$ dispersion, damping rate, and residue of the Higgs mode as a function of $\mu/\gap$, neglecting the influence of the Coulomb interaction. In Sec. \ref{sec:neutral_arbitrary_q}, we extend this analysis and follow the dispersion of the Higgs mode as a function of $\bq$, for arbitrary $\bq$. In Sec. \ref{sec:negative_mu}, we discuss the fate of the Higgs mode for $\mu < 0$. In Sec. \ref{sec:coulomb}, we repeat the analysis of the Higgs mode dispersion, including the effect of the Coulomb interaction. In Sec. \ref{sec:discussion}, we summarize our results.

\section*{\uppercase{Results}}

\subsection{Mean-Field theory}
\label{sec:mean_field}
To analyze the gap function and its fluctuations about equilibrium, we use the functional integral approach \cite{Engelbrecht1997,Diener2008,Pimenov2022}; identical equations can also be obtained diagrammatically \cite{Combescot2006}. Assuming that fermions attract each other via a contact interaction $U(\bx-\by) = -g\delta(\bx-\by)$ and neglecting for the moment the Coulomb interaction, the partition function is given by $Z = \int D(\Bar{\psi}\psi)\exp(-S[\Bar{\psi},\psi])$, where the action $S[\Bar{\psi},\psi]$ in momentum space is given by

\begin{multline}
    S[\bpsi,\psi] = \sum_k \bpsi_\sigma(k)\left(-i\om_n+
    \frac{k^2}{2m} - \mu \right) \psi_\sigma(k)\\
    -
    g\frac{T}{L^2}\sum_{kqp}\bpsi_\up(k+q/2)\bpsi_\down(-k+q/2)\psi_\down(-p+q/2)\psi_\up(p+q/2).
    \label{eq:action_neutral}
\end{multline}

Here, $L^2$ denotes the area of our two-dimensional system, $\mu$ the chemical potential, $\psi_\sigma$ and $\Bar{\psi}_\sigma$ the Grassmann fields describing the fermionic degrees of freedom.
The 3-vectors $k$, $p$, and $q$ label both Matsubara frequency and momentum, e.g. $k = (\om_m,\bk)$. The Matsubara frequencies of $k$ and $p$ are fermionic ($\om_m = (2m+1)\pi T$), while the Matsubara frequency of $q$ is bosonic ($\Om_m = 2\pi m T$). To decouple the quartic interaction, we perform the Hubbard-Stratonovich transformation:
we introduce the complex, bosonic field $\gap_q$, which couples to the $\Bar{\psi}\Bar{\psi}$ terms, and integrate out the fermionic fields $\psi$ and $\Bar{\psi}$. The partition function is then given by a functional integral over the complex field $\gap_q$, $Z = \int D(\gap^*\gap)\exp(-S_\mathrm{eff}[\gap^*,\gap])$, where the effective action is

\begin{equation}
    S_\mathrm{eff}[\gap^*,\gap]
    =
    \frac{\beta L^2 }{g}\sum_{q}\abs{\gap_{q}}^2
    -\mathrm{Tr}\log(-\beta \mathcal{G}^{-1}),
\end{equation}
and the Nambu-Gorkov Green's function in momentum-space is given by

\begin{equation}
    \mathcal{G}^{-1}_{kp} =
    \begin{pmatrix}
    (i\om_n-\xi(\bk))\delta_{kp} & \gap_{k-p}\\
    \gap^*_{p-k} & (i\om_n + \xi(\bk))\delta_{kp}
    \end{pmatrix}.
\end{equation}
where $\xi(\bk) = k^2/2m - \mu$, and $\mu$ is the chemical potential in a superconductor, which at this stage is a parameter.

Thus far, this procedure has been formally exact. To make further progress, we assume that the gap function at equilibrium is spatially uniform and frequency-independent, $\gap_q = \gap \delta_{q,0}$. This solution can be obtained by searching for a saddle point of the effective action. The condition $\delta S_\mathrm{eff}/\delta \gap = 0$ yields the conventional gap equation:

\begin{equation}
    \frac{1}{g} = \int\frac{d^2p}{(2\pi)^2}\frac{\tanh(\beta E_p/2)}{2E_p}.
    \label{eq:gap}
\end{equation}
Here, $E_p = \sqrt{\xi(\bp)^2 + \abs{\gap}^2}$. To handle the UV-divergence on the right-hand side, we impose a high-energy cutoff $\Lambda$, only considering momenta with $p^2/2m < \Lambda$.

The chemical potential $\mu$ is determined by the conservation of particle number. This constraint is enforced by using $n = -\partial \Omega/\partial \mu$, where $\Omega$ is the thermodynamic potential. Within the mean-field approximation, we have in equilibrium, $\Omega = T S_\mathrm{eff}[\abs{\gap}]$, and the equation enforcing particle-number conservation becomes
\begin{equation}
    n = \int \frac{d^2p}{(2\pi)^2} \left[1-\frac{\xi(\bp)}{E_p}\tanh(\beta E_p/2)\right].
    \label{eq:particle_number}
\end{equation}
Due to the U(1) symmetry of the problem, we have the freedom to choose the phase of the order parameter. As such, we henceforth take $\gap$ to be real.

At $T=0$, Eqs. (\ref{eq:gap}) and (\ref{eq:particle_number}) can be solved for $\mu$ and $\gap$ \cite{Randeria1989,Chubukov2016}, and one finds

\begin{align}
    \mu &= E_F - E_0,\\
    \gap &= 2 \sqrt{E_F E_0},
\end{align}

where $E_0 = \Lambda e^{-2/N_0 g}$ is half the binding energy of two fermions in vacuum.

\subsection{Gaussian Fluctuations}
\label{sec:gaussian_fluctuations}
To account for the effects of fluctuations in the order parameter, we expand the gap about the mean-field solution, $\gap(x) = \gap(1+\lambda(x))e^{i\theta(x)} \approx \gap(1 + \lambda(x) + i \theta(x))$, where $\lambda(x)$ and $\theta(x)$ are real dimensionless fields denoting the amplitude and phase fluctuations of the gap, respectively. By inserting this into the effective action and expanding about the saddle point, we find that $S_\mathrm{eff}$ is given to quadratic order by

\begin{equation}
    S_\mathrm{eff} = S_0[\gap]\\
    +\beta L^2
    \gap^2
    \sum_q
    \begin{pmatrix}
    \theta^*_{q} & \lambda_q^*
    \end{pmatrix}
    {\hat M} (i\Om_m,\bq)
    \begin{pmatrix}
    \theta_{q}\\
    \lambda_q
    \end{pmatrix}.
    \label{eq:phase_amp}
\end{equation}

The matrix ${\hat M} (i\Om_m,\bq)$ is the inverse susceptibility for phase and amplitude fluctuations, and its matrix elements are given by

\begin{align}
    M_{++}(z,\bq)
    &= \frac{1}{g} + \frac{1}{2}\chi_{22}(z,\bq)\\
    \label{eq:Mpp}
    M_{--}(z,\bq)
    &= \frac{1}{g} + \frac{1}{2}\chi_{11}(z,\bq)\\
    \label{eq:Mmm}
    M_{+-}(z,\bq)
    &= \frac{1}{2}\chi_{12}(z,\bq)\\
    M_{-+}(z,\bq)
    &=
    -M_{+-}(z,\bq).
\end{align}
The functions $\chi_{ij}(i\Om_m,\bq)$ are defined as $\chi_{ij}(i\Om_m,\bq) = T/L^2 \sum_{\om_m,\bp} \mathrm{Tr}[\mathcal{G}_\mathrm{MF}(i\om_m-i\Om_m/2,\bp-\bq/2)\sigma_i\mathcal{G}_\mathrm{MF}(i\om_m+i\Om_m/2,\bp+\bq/2)\sigma_j]$, where $\sigma_i$ are the Pauli matrices, and $\mathcal{G}_\mathrm{MF}$ is the mean-field Green's function. After performing the Matsubara summation over $\om_m$ and analytically continuing $i\Om_m \rightarrow z$ to complex frequencies in the upper half-plane, we find at $T=0$

\begin{align}
    \chi_{11}(z,\bq)
    &=
    \int \frac{d^2p}{(2\pi)^2}
    \frac{E_++E_-}{E_+E_-}\cdot
    \frac{\xi_+\xi_-+E_+E_--\gap^2}{z^2-(E_++E_-)^2}
    \label{chi_11}\\
    \chi_{22}(z,\bq)
    &=
    \int \frac{d^2p}{(2\pi)^2}
    \frac{E_++E_-}{E_+E_-}\cdot
    \frac{\xi_+\xi_-+E_+E_-+\gap^2}{z^2-(E_++E_-)^2}
    \label{chi_22}\\
    \chi_{12}(z,\bq)
    &=
    -iz\int \frac{d^2p}{(2\pi)^2}
    \frac{1}{E_+E_-}
    \cdot
    \frac{\xi_+E_-+\xi_-E_+}{z^2-(E_++E_-)^2}.
    \label{chi_12}
\end{align}
Here, $\xi_\pm = \frac{(\bp \pm \bq/2)^2}{2m}-\mu$ and $E_\pm = \sqrt{\xi_\pm^2+\gap^2}$. The Higgs susceptibility $\chi_H(z,\bq)$ is
 given by
\begin{equation}
\chi_H(z,\bq) = \frac{M_{++}(z,\bq)}{\det \hat{M}(z,\bq)}.
\label{aa}
\end{equation}
In the high-density limit, one has particle-hole symmetry, so that the off-diagonal matrix elements $M_{+-} = M_{-+} = 0$. In this case, the phase and amplitude fluctuations are completely decoupled, and the Higgs susceptibility is simply given by $\chi_H(z,\bq) \equiv \chi_{--}(z,\bq) = 1/M_{--}(z,\bq)$. Away from the high-density limit, the amplitude-phase coupling is nonzero, and one should use Eq. (\ref{aa}). As discussed in the introduction, we search for the Higgs mode by calculating the location of the poles of $\chi_H^\down(z,\bq)$, the analytical continuation of $\chi_H (z,\bq)$ into the lower half-plane through the real axis at $\omega > 2\Delta$. We search for poles $z_\bq$ of $\chi_H^\down(\om,\bq)$ by solving
\begin{equation}
    \det \hat{M}^\down(z_\bq,\bq) = 0.
\end{equation}

\subsubsection{Analytical Continuation Procedure}
\label{sec:ac_overview}
We now outline how we analytically continue the matrix elements $M_{\sigma\sigma'}(z,\bq)$. In the introduction, we framed analytic continuation as stitching together functions evaluated on different Riemann sheets. Here, we discuss how this procedure is performed computationally.

To this end, recall that the purpose of analytic continuation is to obtain a function which is equal to $M_{\sigma\sigma'}(z,\bq)$ in the upper half-plane, and analytic across the portion of the real axis where $\om > 2\gap$. For this purpose, we define the spectral densities

\begin{equation}
   \rho_{\sigma\sigma'}(\om,\bq)
   =
   \frac{ M_{\sigma\sigma'}(\om+i\delta,\bq) - M_{\sigma\sigma'}(\om-i\delta,\bq)}{-2\pi i}.
    \label{eq:discontinuity_cut}
\end{equation}

With this definition, we trivially have $M_{\sigma\sigma'}(\om-i\delta,\bq) -2\pi i \rho_{\sigma\sigma'}(\om,\bq)=M_{\sigma\sigma'}(\om+i\delta,\bq)$. If we view the expression $M_{\sigma\sigma'}(\om-i\delta,\bq) -2\pi i \rho_{\sigma\sigma'}(\om,\bq)$ as the value of a complex function $M^\down_{\sigma\sigma'}(z,\bq)$ just below the real axis, then we have $M^\down_{\sigma\sigma'}(\om-i\delta,\bq) = M_{\sigma\sigma'}(\om+i\delta,\bq)$. Using this, the following function is by construction analytic across the real axis:

\begin{equation}
    M_{\sigma\sigma'}^\down(z,\bq)
    =
    \begin{cases}
    M_{\sigma\sigma'}(z,\bq) & \mathrm{Im}(z) > 0,\\
    M_{\sigma\sigma'}(z,\bq)-2\pi i \rho_{\sigma\sigma'}(z,\bq)
    &\mathrm{Im}(z) <0.
    \end{cases}
    \label{eq:M_ac}
\end{equation}

\begin{figure}
\includegraphics[width=\columnwidth]{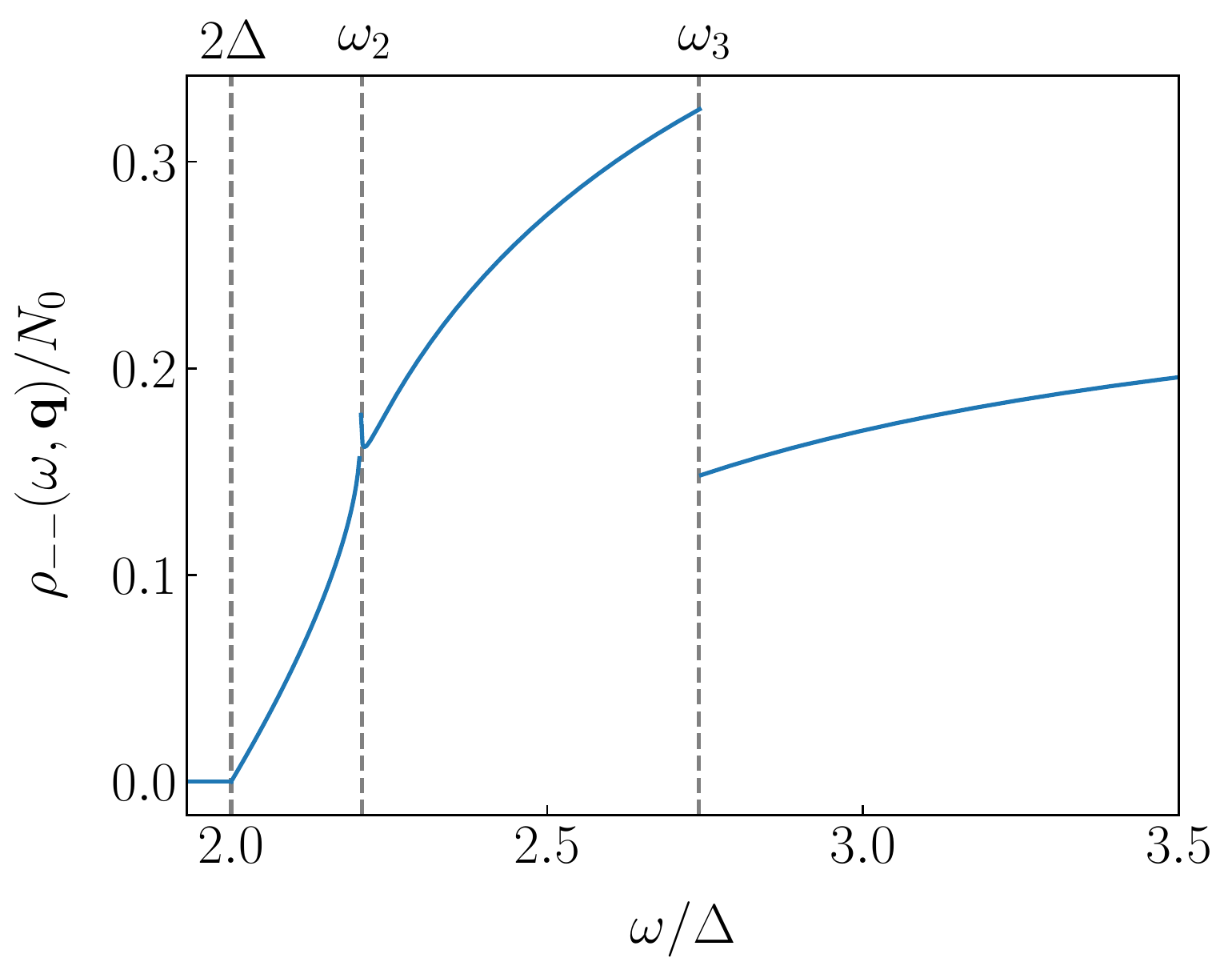}
\caption{\label{fig:rho}  The spectral density $\rho_{--}(\om,\bq)$
at $\mu = \gap$ and $q = 0.5\sqrt{2m\mu}$, as a function of $\om$. The frequencies $\omega_2$  at which $\rho_{--}(\om,\bq)$ has a kink, and $\omega_3$, at which it is discontinuous, are defined in the text.}
\end{figure}

Note that in this equation, we have replaced $\rho_{\sigma\sigma'}(\om,\bq)$ with its analytical continuation away from the real axis, $\rho_{\sigma\sigma'}(z,\bq)$. This requires care, since $\rho_{\sigma\sigma'}(\om,\bq)$ is not analytic for all $\omega > 2 \Delta$ -- it has a kink at some higher frequency $\omega_2$, and a discontinuity at even higher $\omega_3$. We illustrate this in Fig. \ref{fig:rho}, where we plot $\rho_{--}(\om,\bq)$ as a function of $\om$ for $\mu = \gap$ and $q = 0.5\sqrt{2m\mu}$. These kinks and discontinuities result from Lifshitz transitions, which we discuss in Sec. \ref{app:analytic_continuation_discussion} of the SI. To obtain a function which we can analytically continue away from the real axis, we must restrict the domain of $\rho_{--}(\om,\bq)$ to a subset of the real axis on which $\rho_{--}(\om,\bq)$ is analytic. Similar consideration holds for the other spectral densities $\rho_{\sigma\sigma'}(\om,\bq)$.

Once we choose a region of the real axis on which $\rho_{
\sigma\sigma'}(\om,\bq)$ is analytic, we analytically continue $\rho_{\sigma\sigma'}(\om,\bq)$ to obtain the complex function $\rho_{\sigma\sigma'}(z,\bq)$, and use Eq. (\ref{eq:M_ac}) to obtain the analytic continuation of $M_{\sigma\sigma'}(z,\bq)$ into the lower half-plane.
 The resulting $M^\down_{\sigma\sigma'}(z,\bq)$ is analytic across the region of the real axis we have chosen.

Different choices of domains for $\rho_{\sigma\sigma'}(\om,\bq)$ lead to distinct analytic behaviors in $M^\down_{\sigma\sigma'}(z,\bq)$, and corresponds to defining $M^\down_{\sigma\sigma'}(z,\bq)$ using different unphysical Riemann sheets in the lower half-plane.
Expressions for the spectral densities for arbitrary $\bq$ and $\om$ can be found in Sec. \ref{app:analytic_continuation_full} of the SI, as well as their analytic continuations through the different regions of $\om$. For general values of $\bq$, $\mu$, and $\om$, the analytically continued matrix elements contain hyper-elliptic integrals, which we handle numerically.  However, in some limits, expressions for the analytically continued matrix elements $M^\down_{\sigma\sigma'}(z,\bq)$ turn out to be relatively simple (see the next section.)

Since we expect the Higgs mode to have frequencies just above the boundary of the two-particle continuum, we analytically continue $\rho_{\sigma\sigma'}(\om,\bq)$ (and hence $M_{\sigma\sigma'}(\om,\bq)$) through the region $(2\gap, \om_2)$ for $\mu > 0$. Doing so leads to matrix elements which are analytic across $(2\gap,\om_2)$, but discontinuous across other regions of the real axis. When $\mu < 0$, the lower bound of the two-particle continuum is instead at $\om_3$, which becomes the lower bound of the
%particle-hole
two-particle
continuum $\om_\mathrm{min}$,
 so we analytically continue the matrix elements through the region $(\om_\mathrm{min},\infty)$. This procedure leads to matrix elements which are analytic across
$(\om_\mathrm{min},\infty)$
, but discontinuous across, e.g. the region
$(0,\om_\mathrm{min})$.

\subsection{The long-wavelength dispersion of the Higgs mode}
\label{sec:neutral_analytic}
In this section, we calculate the dispersion of the Higgs mode at small $\bq$ and $\mu > 0$.
As discussed in the previous section, this is done by first constructing $\chi_H^\down(z,\bq)$, the analytical continuation of $\chi_H(z,\bq)$ into the lower half-plane through $\om \in (2\gap,\om_2)$. This region is chosen because we expect the Higgs mode to begin at $z_{\bq=0}=2\Delta$ and disperse quadratically with $q$ to larger values of $\Re(z_\bq)$. We then search for a pole in $\chi^\down_H(z,\bq)$ of the form $z_\bq = 2\Delta + \zeta \frac{q^2}{2m} \frac{\mu}{\Delta}$. We will see that $\om_2 = 2\Delta +\frac{q^2}{2m} \frac{\mu}{\Delta}$ at small $\bq$. Accordingly, we constrain $\Re(\zeta)$ to take values in the interval $(0,1)$-- this ensures $\Re(z_\bq) \in (2\gap,\om_2)$.

Below we compute the matrix elements $M_{\sigma\sigma'}(z,\bq)$ in the upper half-plane and analytically continue them one by one through the real axis. We begin by calculating $M_{--}(z,\bq)$ using Eq. (\ref{eq:Mmm}). Combining Eq. (\ref{chi_22}) with the gap equation $\frac{1}{g} = \frac{1}{2}\int \frac{d^2p}{(2\pi)^2}\frac{1}{E}$, we express $M_{--}(z,\bq)$ as
\begin{multline}
    M_{--}(z,\bq)
    =\\
    \frac{1}{4}\int \frac{d^2p}{(2\pi)^2}
    \frac{E_++E_-}{E_+E_-}\cdot
    \frac{z^2-4\gap^2-(\xi_+-\xi_-)^2}{z^2-(E_++E_-)^2}.
\end{multline}
Evaluating this integral (technical details can be found in Sec. \ref{app:technical_details} of the SI), we obtain
\begin{equation}
    M_{--}(z_\bq,\bq)
    =
    -i N_0 \frac{v_\mu q}{2\gap}\sqrt{\zeta}E(\frac{1}{\sqrt{\zeta}}),
\end{equation}
where $N_0 = m/2\pi$ is the density of states per spin in 2D, and $E(z)$ is the complete elliptic integral of the second kind \footnote{To be explicit, here we use the convention where the complete elliptic integrals of first and second kind are defined as
$K(z) = \int_0^{\pi/2} dx/\sqrt{1-z^2 \cos^2{x}}$ and $E(z) = \int_0^{\pi/2} dx \sqrt{1-z^2 \cos^2{x}}$ (see Eq. 19.2.8 of Ref. \cite{NIST:DLMF}.) This is different from the convention used in Mathematica, where $z^2$ in the integrands of $K(z)$ and $E(z)$ are replaced by $z$.}.

We now analytically continue $M_{--}(z_\bq,\bq)$ into the lower half-plane of complex $z$ across $\omega \in (2\gap, \omega_2)$, i.e., across $\zeta \in (0,1)$. This is achieved by substituting  $E(\frac{1}{\sqrt{\zeta}})$ at $\Im(\zeta)=0^+$ by $E(\frac{1}{\sqrt{\zeta}})+2i(E(\sqrt{1-\zeta^{-1}})-K(\sqrt{1-\zeta^{-1}}))$ at $\Im(\zeta) =0^-$ (see Sec. \ref{app:analytic_continuation_EK} of the SI for a proof). The analytic continuation of $M_{--}(z_\bq,\bq)$ is therefore given by

\begin{multline}
    M_{--}^\down(z_\bq,\bq)
    =\\
    \begin{cases}
    -i N_0 \frac{v_\mu q}{2\gap}\sqrt{\zeta} E\left(\frac{1}{\sqrt{\zeta}}\right), & \Im(\zeta) > 0,\\
    -iN_0\frac{v_\mu q}{2\gap}\sqrt{\zeta}
    \big[
    E(\frac{1}{\sqrt{\zeta}})\\
    +2i(E(\sqrt{1-\zeta^{-1}})-K(\sqrt{1-\zeta^{-1}}))
    \big], & \Im(\zeta) < 0.\\
    \end{cases}
\end{multline}

We use the same tactics to compute  $M_{++}(z_\bq,\bq)$, given by
\begin{multline}
    M_{++}(z,\bq)
    =\\
    \frac{1}{2}\int \frac{d^2p}{(2\pi)^2}\left(
    \frac{E_++E_-}{E_+E_-}\cdot
    \frac{\xi_+\xi_-+E_+E_-+\gap^2}{z^2-(E_++E_-)^2}
    +\frac{1}{E}
    \right).
\end{multline}
Evaluating the momentum integral in the same way as for $M_{--}(z_\bq,\bq)$, we obtain

\begin{equation}
    M_{++}(z_\bq,\bq)
    =-iN_0\frac{2\gap}{v_\mu q}\frac{1}{\sqrt{\zeta}}
    K\left(\frac{1}{\sqrt{\zeta}}\right),
\end{equation}
where $K(z)$ is the complete elliptic integral of the first kind. The analytical continuation through the interval of the real axis where $\zeta \in (0,1)$ is achieved by substituting  $K(\frac{1}{\sqrt{\zeta}})$ at $\Im(\zeta) =0^+$ by $K(\frac{1}{\sqrt{\zeta}}) - 2iK(\sqrt{1-\zeta})\sqrt{\zeta}$  when $\Im(\zeta) < 0$ (see Sec. \ref{app:analytic_continuation_EK} of the SI). We then obtain

\begin{multline}
    M_{++}^\down(z_\bq,\bq)
    =\\
    \begin{cases}
    -iN_0\frac{2\gap}{v_\mu q}\frac{1}{\sqrt{\zeta}}
    K\left(\frac{1}{\sqrt{\zeta}}\right), & \Im(\zeta) > 0\\
    -iN_0\frac{2\gap}{v_\mu q}\frac{1}{\sqrt{\zeta}}
    \left[
    K\left(\frac{1}{\sqrt{\zeta}}\right) - 2iK(\sqrt{1-\zeta})\sqrt{\zeta}\right], & \Im(\zeta) < 0.\\
    \end{cases}
\end{multline}

We now turn to the matrix elements $M_{+-}(z,\bq)$ and $M_{-+}(z,\bq)$, which couple amplitude and phase fluctuations. Since $M_{-+}(z,\bq)=-M_{+-}(z,\bq)$, we focus on $M_{+-}(z,\bq)$. We recall that in the end, we need to solve $\det \hat{M}^\down(z_\bq,\bq)  = M^\down_{++}(z_\bq,\bq)M^\down_{--}(z_\bq,\bq) - M_{+-}^\down(z_\bq,\bq)M_{-+}^\down(z_\bq,\bq)=0$. At small $\bq$ we have  $M_{++}(z_\bq,\bq) = O(1/q)$ and $M_{--}(z_\bq,\bq) =O(q)$. Since their product is $O(1)$, it is sufficient to compute $M_{+-}(
z_\bq,\bq)$ at $q=0$, where $z_{\bq=0}=2\gap$. The matrix element $M_{+-}(2\gap,0)$ is purely imaginary and is given by

\begin{align}
    M_{+-}(2\gap,0)
    &=
    i\frac{\gap}{2}N_0\int_{-\mu}^\infty \frac{d\xi}{\xi E} \label{eq:undef}\\
    &=
    i \frac{N_0}{4}\log(\frac{\sqrt{\mu^2+\gap^2}+\gap}{\sqrt{\mu^2+\gap^2}-\gap}).
\end{align}

\subsubsection{High-Density Limit}
\label{sec:high_density}
In the high-density limit where $\mu \approx E_F \gg \gap$, the amplitude-phase coupling arising from $M_{+-}$ is
 small in $\Delta/\mu$ and can be neglected. The Higgs susceptibility $\chi^\down _H(z,\bq)$ then reduces to $\chi^\down_H(z,\bq) = 1/M^\down_{--}(z,\bq)$. The parameter $\zeta$, which determines the location of the Higgs mode in the lower half-plane is the solution of $M^\down_{--}(\zeta) = 0$, i.e., of
\begin{equation}
    E(\frac{1}{\sqrt{\zeta}})
    +2i(E(\sqrt{1-\zeta^{-1}})-K(\sqrt{1-\zeta^{-1}}))
    =
    0.
\end{equation}

The solution of this transcendental equation is $\zeta = 0.5-0.4308i$, so that the location of the Higgs mode is

\begin{equation}
    z_\bq = 2\gap + (0.5-0.4308i)\frac{q^2}{2m} \frac{\mu}{\Delta}.
\end{equation}
The dispersion of the Higgs mode at small $\bq$ is given by $\omega_\bq=\Re(z_\bq) =2\gap + \frac{q^2}{4m} \frac{\mu}{\Delta}$. The damping rate of the Higgs mode, $\gamma_\bq$, is quadratic in $\bq$, as in 3D \cite{Andrianov1976,Kurkjian2019}.

\subsubsection{Away from the high-density limit}
\label{sec:arbitrary_mu}
Away from the high-density limit, $M_{+-}$ has to be kept. The susceptibility $\chi^\down_H (z_\bq,\bq)$ has the form

\begin{widetext}
\begin{multline}
    \chi^\down_H(z_\bq,\bq)
    =
    \frac{i}{N_0} \frac{2\gap}{v_\mu q}\frac{1}{\sqrt{\zeta}}\\
    \times
    \frac{K(\frac{1}{\sqrt{\zeta}})-2iK(\sqrt{1-\zeta})\sqrt{\zeta}}
    {\left[K(\frac{1}{\sqrt{\zeta}}) - 2iK(\sqrt{1-\zeta})\sqrt{\zeta}\right]
    \left[
    E(\frac{1}{\sqrt{\zeta}})+2i\left(E(\sqrt{1-\zeta^{-1}})-K(\sqrt{1-\zeta^{-1}})\right)
    \right]
    +
    \frac{1}{16}\left(\log(\frac{\sqrt{\mu^2+\gap^2}+\gap}{\sqrt{\mu^2+\gap^2}-\gap})\right)^2}.
    \label{eq:chi_mm}
\end{multline}
\end{widetext}

The position of the Higgs mode is determined by the condition

\begin{widetext}
\begin{equation}
    \left[K(\frac{1}{\sqrt{\zeta}}) - 2iK(\sqrt{1-\zeta})\sqrt{\zeta}\right]
    \left[
    E(\frac{1}{\sqrt{\zeta}})+2i\left(E(\sqrt{1-\zeta^{-1}})-K(\sqrt{1-\zeta^{-1}})\right)
    \right]
    +
    \frac{1}{16}\left(\log(\frac{\sqrt{\mu^2+\gap^2}+\gap}{\sqrt{\mu^2+\gap^2}-\gap})\right)^2 = 0.
    \label{eq:zeta}
\end{equation}
\end{widetext}

\begin{figure}
\includegraphics[width=\columnwidth]{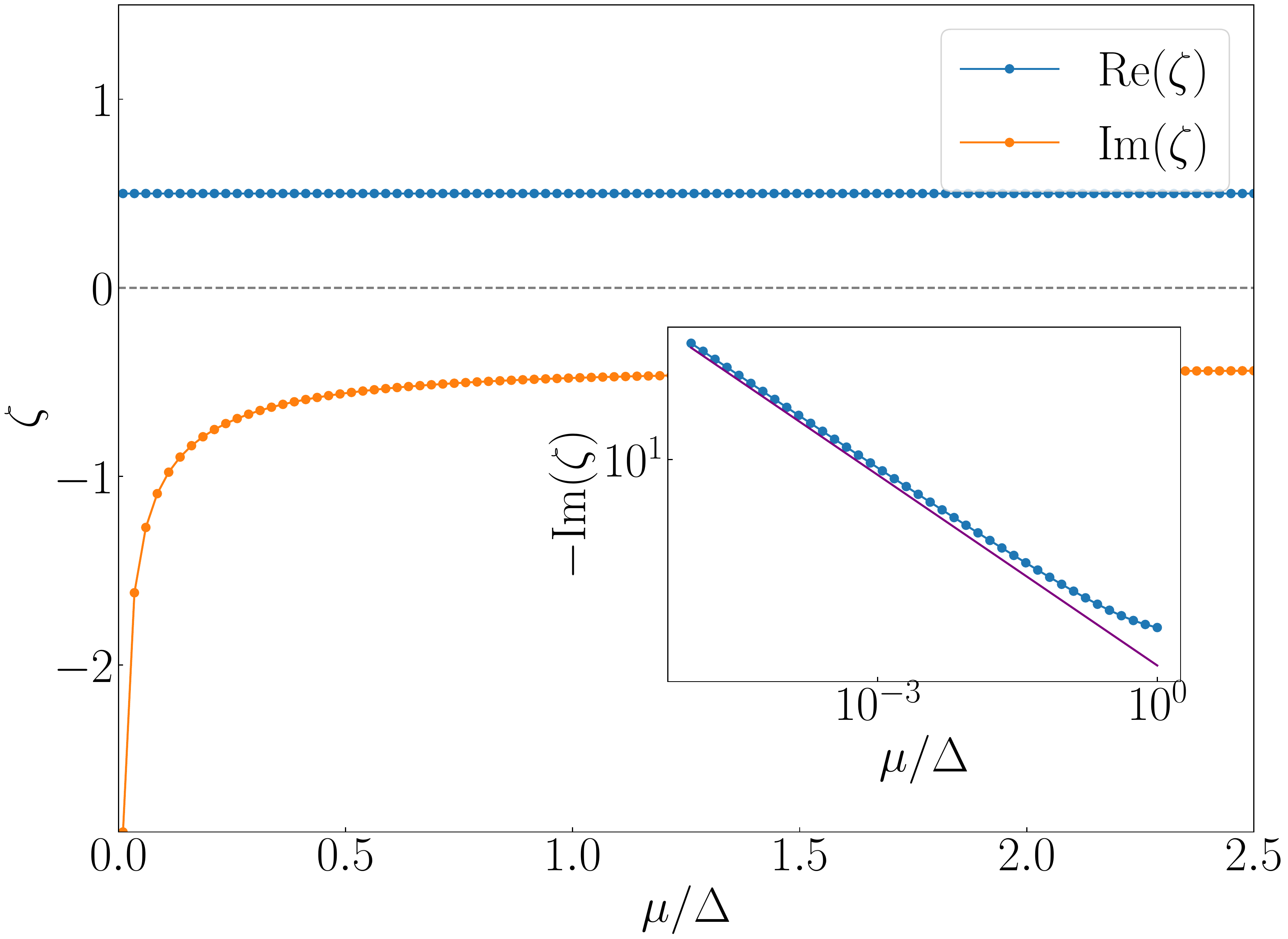}
\caption{\label{fig:zeta} The real and imaginary parts of $\zeta$ given by the solution of Eq. (\ref{eq:zeta}), as a function of $\mu$. The quantity $\zeta$ is related to the dispersion of the Higgs mode as $\om_\bq = 2\gap + \zeta \frac{q^2}{2m}\frac{\mu}{\gap}$. We see that $\Re(\zeta)=0.5$ for all $\mu/\gap$, while $\Im(\zeta)$
diverges at small $\mu$. The inset highlights that the divergence is a power-law. The purple line in the inset is the analytical expression for $\Im(\zeta)$, Eq. (\protect\ref{uu}).}
\end{figure}

We numerically solve this equation for $\zeta$ for any value of $\gap$ and $\mu > 0$, where this equation is valid. We present the results in Fig. \ref{fig:zeta}. We see that $\zeta$ evolves as a function of $\mu$, but, remarkably, $\Re(\zeta) = 0.5$ for \textit{all} values of $\mu$. With this, the dispersion at small $\bq$ is given for all $\mu$ by $\omega(\bq) =2\gap + \frac{q^2}{4m} \frac{\mu}{\Delta}$.

The fact that $\Re(\zeta) = 0.5$ holds for all $\mu>0$ follows from a special reflection symmetry of the equation for the pole location in 2D. We show in Sec. \ref{app:reflection_proof} of the SI that if $\zeta$ is a solution to Eq. (\ref{eq:zeta}), its reflection across the line where $\Re(\zeta)=0.5$ is also a solution. Combining this with the fact that Eq. (\ref{eq:zeta}) has a unique solution, we immediately find that $\Re(\zeta)$ must equal $0.5$ for all values of $\mu$. We see therefore that $\Re(\zeta)$ remains inside the interval $(0,1)$ for any positive value of $\mu$. This is in contrast to the behavior in 3D, where $\Re(\zeta)$ becomes negative for $\mu < \mu_c = 0.8267\gap$ \cite{Kurkjian2019}.

The damping rate of the Higgs mode increases with decreasing $\mu$ and diverges in the limit where $\mu \ll \gap$ as
\begin{equation}
    \Im(\zeta) \approx -i \frac{e}{16}\sqrt{\frac{2\gap}{\mu}}.
    \label{uu}
\end{equation}
In the inset of Fig. \ref{fig:zeta}, we overlay this expression on the numerical solution of Eq. (\ref{eq:zeta}), finding good agreement for smaller values of $\mu/\gap$.

\subsubsection{The Residue of the Higgs mode}
The residue of the Higgs mode is defined as $Z_\bq = \lim_{z\rightarrow z_\bq}(z-z_\bq)\chi^\down_H(z,\bq)$. We find
\begin{widetext}
\begin{equation}
    Z_\bq
    = i\frac{v_\mu q}{2N_0}
    \frac{\frac{1}{\sqrt{\zeta}}K(\frac{1}{\sqrt{\zeta}}) - 2iK(\sqrt{1-\zeta})}
    {\frac{d}{d\zeta}\left[K(\frac{1}{\sqrt{\zeta}}) - 2iK(\sqrt{1-\zeta})\sqrt{\zeta}\right]
    \left[
    E(\frac{1}{\sqrt{\zeta}})+2i(E(\sqrt{1-\zeta^{-1}})-K(\sqrt{1-\zeta^{-1}}))
    \right]},
    \label{eq:residue}
\end{equation}
\end{widetext}
where $\zeta$ is the solution of Eq. (\ref{eq:zeta}).

\begin{figure}
\includegraphics[width=\columnwidth]{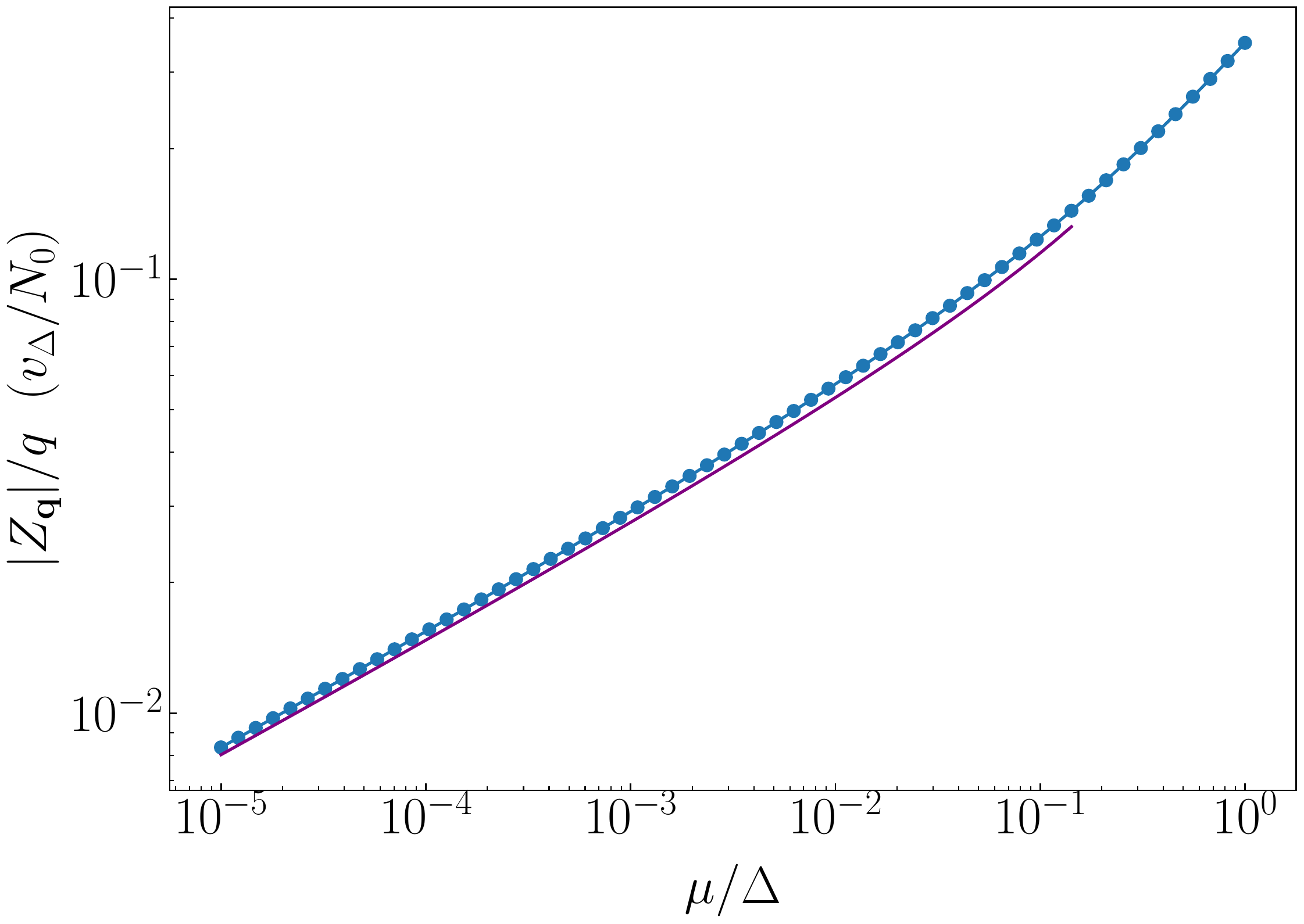}
\caption{\label{fig:residue} The ratio $\abs{Z_\bq}/q$ as a function of $\mu/\gap$, Eq. (\ref{eq:residue}). The purple line-- an approximate analytic expression at small $\mu/\Delta$: $\abs{Z_\bq}/q \propto (\mu/2\gap)^{1/4}$,  Eq. (\ref{eq:res_approx}).}
\end{figure}

As expected, $Z_\bq$ goes to zero in the long-wavelength limit. This reflects the disappearance of the pole in the susceptibility $\chi_H(z,\bq)$ at $\bq = 0$. At high density,
$Z_\bq = -0.2474(1-i)v_F q/N_0$. In the opposite limit where $\mu/\gap \rightarrow 0$,
\begin{equation}
    Z_\bq \approx -\frac{(1-i)e^{1/2}}{16}\frac{v_\gap q}{N_0} \left(\frac{\mu}{2\gap}\right)^{1/4}
    \label{eq:res_approx}
\end{equation}
up to logarithmic corrections. Here, $v_\gap$ is defined through $mv_\gap^2/2 = \gap$. From this expression, we see that the residue of the pole goes to zero at small $\mu$ as $(\mu/\gap)^{1/4}$. In Fig. \ref{fig:residue}, we plot $\abs{Z_\bq}/q$ at small $\bq$, using both the exact expression of Eq. (\ref{eq:residue}) and the approximate expression of Eq. (\ref{eq:res_approx}), including logarithmic corrections to Eq. (\ref{eq:res_approx}). From Fig. \ref{fig:residue}, we see that there is good agreement between the exact and approximate expressions for $Z_\bq$ at small $\mu/\gap$.

\subsubsection{Susceptibility $\chi_H(\omega,\bq)$ along the real axis
}
To calculate the observable $\Im\chi_H(\omega +i\delta,\bq)$, we recall that for $\zeta \in (0,1)$, $\chi_H(\omega +i\delta,\bq) =\chi^\down_H(\omega,\bq)$. Near the location of the pole in the lower half-plane, we can write $\chi^\down_H(z,\bq) = A + Z_\bq/(z-\om_\bq+i\gamma_\bq)$. If the pole is close to the real axis, we then expect the spectral function $\Im \chi_H(\om +i\delta, \bq)$ to be approximately given by $\Im(A) + \Im(Z_\bq/(\om-\om_\bq+i\gamma_\bq))$. This has a peak at $\omega = \om_\bq$, and an approximate width of $\gamma_\bq$.

\begin{figure*}
\centering
\includegraphics[width=0.95\textwidth]{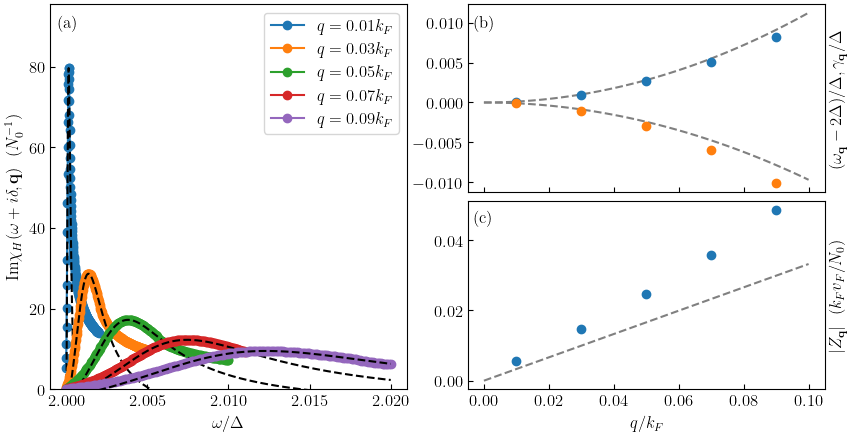}
\caption{\label{fig:fits}
(a) The spectral function $\Im \chi_H(\om+i\delta,\bq)$ as a function of $\om$ for different values of $\bq$ at $E_F = 10E_0$. The dashed black lines are the fits to $\Im \chi_H(\om+i\delta,\bq)= C + \Im(Z_\bq/(\om-\om_\bq+i\gamma_\bq))$. (b-c) The behavior of $\om_\bq$, $\gamma_\bq$, and $\abs{Z_\bq}$ as a function of $\bq$, extracted from the fits in panel (a). The dashed gray curves correspond to our analytical expressions.}
\end{figure*}

In Fig. \ref{fig:fits}(a), we plot the spectral function $\Im\chi_H(\om+i\delta,\bq)$ obtained numerically using Eqs. (\ref{eq:Mpp}-\ref{aa}) for five momenta between $q=0$ and $q=0.1k_F$, using $\delta = 10^{-5}\gap$ and $E_F = 10E_0$ (corresponding to $\mu/\gap \approx 1.42$). The overlaid dashed black lines are the curves obtained by fitting $\Im\chi_H(\om+i\delta,\bq)$ to the function $C+ \Im(Z_\bq/(\om-\om_\bq+i\gamma_\bq))$. Since we expect this functional form to only be meaningful near the resonance of the spectral function, we restrict each fit to only use data points where $\Im\chi_H(\om+i\delta,\bq) > 0.8\max(\Im\chi_H(\om+i\delta,\bq))$.

We see that at small $\bq$, $\Im\chi_H(\om+i\delta,\bq)$ closely resembles the one-sided square-root singularity we expect from $q=0$, albeit with a peak above $2\gap$. With increasing $q$, this peak in the spectral function broadens substantially and moves to larger values of $\om$. In Figs. \ref{fig:fits}(b,c), we present the extracted values of $\om_\bq$, $\gamma_\bq$ and $\abs{Z_\bq}$ from fitting each of the five curves. We have also added a dashed gray curve to denote the results expected from the analytical expressions derived above. We find good agreement in the dispersion $\om_\bq$ and damping rate $\gamma_\bq$, while the agreement between the numerical and analytical results for $\abs{Z_\bq}$ is a bit more ambiguous.

In particular, the values of $\abs{Z_\bq}$, extracted from fitting to the numerical data, consistently lie above the line expected from our analytical results. We attribute this to the ambiguity in the method used to fit the data: although we restrict each fit to only use data points above some threshold, $\Im\chi_H(\om+i\delta,\bq) > 0.8\max(\Im\chi_H(\om+i\delta,\bq))$, this 80\% threshold is rather arbitrary. We find that the values of $\abs{Z_\bq}$ extracted from fitting to the data are rather sensitive to the precise threshold used \footnote{$\om_\bq$ and $\gamma_\bq$ also change with the threshold, but continue to fit the analytical expressions relatively well regardless of the precise threshold used.}. Nonetheless, we find that the values of $\abs{Z_\bq}$ extracted from fitting to the data agree with the analytical results within a factor of $\sim 2$ for all reasonable thresholds. Moreover, we find that for all thresholds employed, (i) $\abs{Z_\bq}$ increases linearly with $q$, and (ii) the phase of $Z_\bq$, i.e. $\arg(Z_\bq)$, is approximately $3\pi/4$. Both behaviors agree with the analytical expressions in Eq. (\ref{eq:residue}) and Eq. (\ref{eq:res_approx}).

We note in passing that our result that the peak in $\Im \chi_H(\om +i\delta, \bq)$ in 2D exists for all $\mu >0$
(in contrast to 3D, where the peak only exists for $\mu > 0.8267 \Delta$), agrees with a previous numerical study, which found that the Higgs mode is more visible in the dynamical structure factor in 2D compared to 3D \cite{Zhao2020}.
We also note that as $\mu \to 0$, the boundary frequency $\omega_2 = 2\Delta + \frac{q^2}{2m} \frac{\mu}{\Delta}$ approaches $2\Delta$, i.e. the interval $(2\Delta, \omega_2)$ vanishes at $\mu =0$. This is in line with the vanishing of the residue of the Higgs mode $Z_\bq \sim (\mu/\gap)^{1/4}$ as $\mu \to 0$.

\subsection{The Higgs mode at larger values of $\bq$}
\label{sec:neutral_arbitrary_q}
\begin{figure*}[t]
\includegraphics[width=\textwidth]{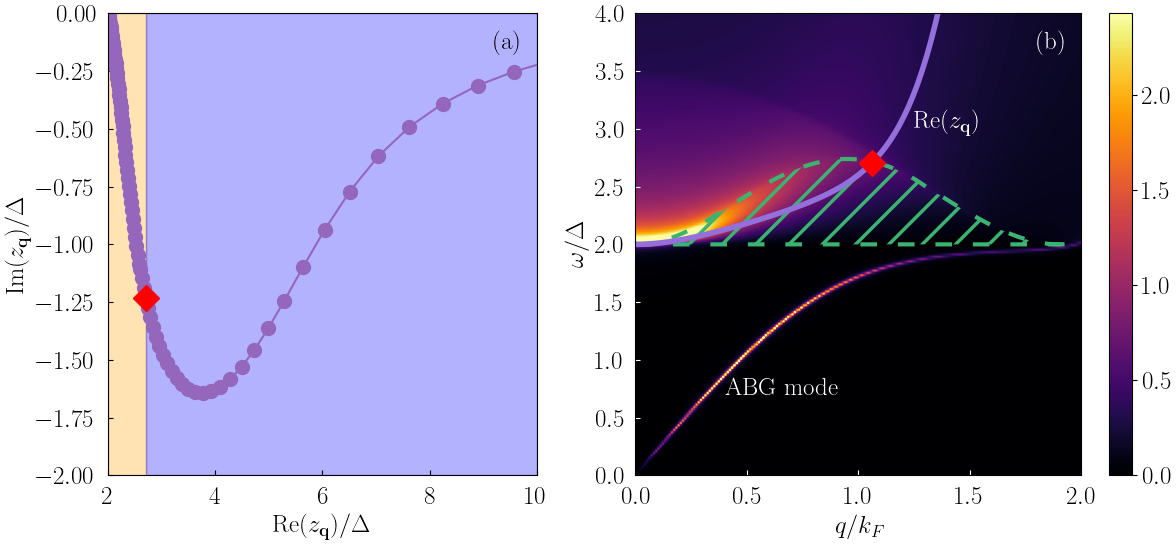}
\caption{\label{fig:chi_Ef_10}
The location of the pole and the spectral function $\Im\chi_H(\om+i\delta,\bq)$ in the BCS regime at $E_F = 10 E_0$
($\mu = 9E_0$, $\gap = 2\sqrt{10}E_0$, and $\mu/\gap= 1.63$).
Panel (a): the path of the pole of $\chi^\down_H(z,\bq)$ through the lower half-plane with increasing $q$. For $q > q_c = 1.08k_F$, the pole is hidden below a branch cut, and does not lead to a resonance in the spectral function $\Im\chi_H(\om+i\delta,\bq)$. The transition point where the pole becomes hidden is marked with a red diamond. We highlight the region where a pole leads to a peak in $\Im \chi_H(\om+i\delta,\bq)$ with a light-orange background, and the region where the pole is hidden below a branch cut with a light-blue background.
(b) The spectral function $\Im \chi_H(\om+i\delta,\bq)$  (color-coding on the right).
Purple
line -- $\Re(z_\bq)$, where $z_\bq$ is the position of the pole of $\chi_H^\down(z,\bq)$. The hatched
green
region corresponds to frequencies between $2\gap$ and $\om_2$. The two frequencies differ by $q^2$ at small $q$ and merge again at $q = 2\sqrt{2m\mu}$. Values of $\Re(z_\bq)$ in the hatched region correspond to those in the light-orange region in panel (a). Outside this region, the pole is hidden. The data show that the peak inside the hatched region rapidly broadens with increasing $q$. The mode below $2\Delta$ in panel (b) is the ABG mode.}
\end{figure*}

So far, we have analyzed the Higgs mode at small $\bq$. In this section, we go beyond the small-$\bq$ regime, continuing to take $\mu > 0$. We find how the Higgs mode evolves as a function of $\bq$ by numerically solving for the position of the pole of $\chi^\down_H(z,\bq)$ without assuming that $q$ is small (see Sec. \ref{app:analytic_continuation_full} of the SI for details.) Our results are shown in Fig. \ref{fig:chi_Ef_10} for $E_F = 10E_0$.

In Fig. \ref{fig:chi_Ef_10}(a), we show how the pole of $\chi^\down_H(z,\bq)$ moves through the lower half-plane as a function of $q$. With increasing $q$, the pole at $z_\bq$ quickly moves away from the real axis, leading to heavier damping of the Higgs mode. As $q$ increases beyond some threshold $q_c$, $\Re(z_\bq)$ becomes larger than $\omega_2$. In this situation, $\chi^\down_H(z,\bq)$ is no longer continuous upon crossing the real axis. At this point, the pole in the lower half-plane becomes a hidden mode-- although the pole exists, it does not lead to a peak in $\Im\chi_H(\om+i\delta,\bq)$ since it lies below a branch cut of $\chi^\down_H(z,\bq)$. To highlight this transition, we mark the point where the Higgs mode becomes hidden with a red diamond. For $q < q_c$, the Higgs mode is observable; we highlight these values via a light-orange background. Similarly, the Higgs mode is hidden for $q > q_c$, and we highlight this region with a light-blue background.

In Fig. \ref{fig:chi_Ef_10}(b), we present the spectral function $\Im\chi_H(\om+i\delta,\bq)$, as well as the dispersion of the Higgs mode, $\Re(z_\bq)$, obtained by numerically solving $\det \hat{M}(z_\bq,\bq) = 0$ for all $\bq$. Additionally, we have added a hatched region corresponding to $(2\gap,\om_2)$-- values of $\Re(z_\bq)$ in this region are not hidden below a branch cut, and lead to a peak in the spectral function. From this plot, we see a sharp bright feature in the spectral function near $z=2\gap$ and $q=0$, which broadens with $q$. The peak in $\Im\chi_H(\om+i\delta,\bq)$ disappears around $q= 1.1k_F$, where $\Re(z_\bq)$ becomes larger than $\omega_2$. This behavior is fully consistent with Fig. \ref{fig:chi_Ef_10}(a), where the pole moves with increasing $q$ deeper into the lower half-plane, and is eventually hidden below a branch cut. As in Fig. \ref{fig:chi_Ef_10}(a), we highlight the moment where the pole becomes hidden with a red diamond. Also visible in Fig. \ref{fig:chi_Ef_10}(b) is the ABG mode, which disperses linearly at small $q$. Its visibility in the Higgs (amplitude) susceptibility arises from the phase-amplitude coupling, which is nonzero at finite $\mu/\gap$.

\subsection{The Higgs mode for $\mu < 0$}
\label{sec:negative_mu}
Thus far, we have restricted ourselves to the case where $\mu > 0$. In this section, we consider the behavior of the Higgs mode for $\mu < 0$. For this analysis, we first note that both $\omega_2$ and $\omega_3$, depicted in Fig. \ref{fig:rho}, approach $2\Delta$ as $\mu$ approaches zero from above. At $\mu =0$, $\omega_2 = \omega_3 =2\Delta$, and the interval $(2\gap,\infty)$ coincides with the interval $(\om_3,\infty)$. A simple analysis shows that for $\mu <0$, the frequency
$\om_3 = 2\sqrt{\gap^2+(\abs{\mu}+q^2/8m)^2}$
 becomes the  lower boundary for the branch cut in $\chi_H(\om,\bq)$, $\om_\mathrm{min}$,
  i.e. a branch cut exists for $\omega > \om_3 = \om_\mathrm{min}$.

This change in the branch cut boundary can also be understood by thinking of $2\gap$, $\om_2$, and $\om_3$ as branch points of the Higgs susceptibility. From this perspective, the branch points $2\gap$ and $\om_2$ annihilate at $\mu = 0$, leaving only the branch point at $\om_3$ for $\mu < 0$. Mathematically, the disappearance of the branch points at $2\gap$ and $\om_2$, as $\mu$ changes sign, corresponds to a change in the topology of the Riemann surface at $\mu = 0$.

To search for a resonance in the susceptibility $\chi_H(\om+i\delta,\bq)$, we now investigate its analytical continuation through the real axis at $\omega > \omega_3$ and  search for a pole in $\chi^\down_H(z,\bq)$ in the lower half-plane. Skipping the details of the calculations, we find that a pole exists at some $z=z_\bq$, but $\Re(z_\bq) < \omega_3$. That is, the pole at $\mu <0$ is hidden, since $\chi^\down_H(z,\bq)$ is discontinuous across the real axis at $\om = \Re(z_\bq)$. We therefore expect that this pole does not lead to a peak in $\Im\chi_H(\om+i\delta,\bq)$.

\begin{figure}
\includegraphics[width=\columnwidth]{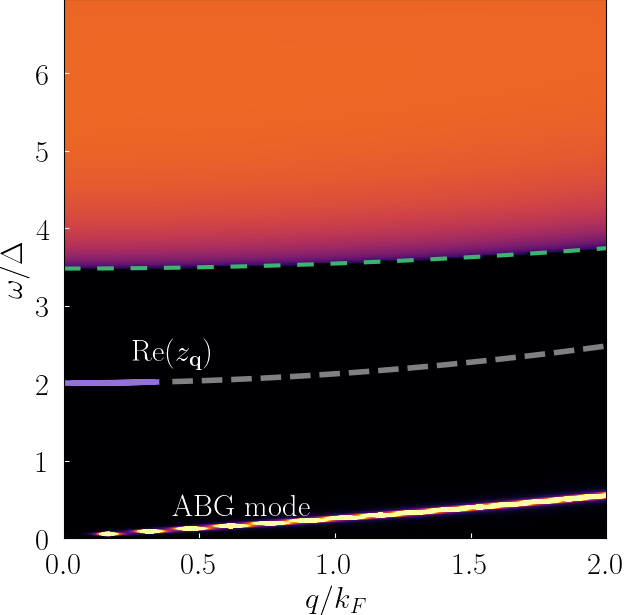}
\caption{$\Im \chi_H(\om+i\delta,\bq)$ in the BEC regime, at $E_F = 0.1E_0$ ($\mu = -0.9E_0$, $\gap = 0.63E_0$, and $\mu/\gap = -1.42$). The color coding is the same as in Fig.\protect\ref{fig:chi_Ef_10}. The dashed
green
curve is the edge of the two-particle continuum, which at finite $q$ is
$\om_\mathrm{min}
= 2\sqrt{\Delta^2 + (\abs{\mu} + q^2/8m)^2}$. The
purple
line shows $\Re(z_\bq)$, where $z_\bq$ is the location of the pole of $\chi^\down_H(z,\bq)$. This Re$z_\bq$ is obtained numerically by solving $\det \hat{M}^\down(z_\bq,\bq) = 0$ for each value of $\bq$. The pole positions at larger $q$ are not shown due to numerical difficulties \cite{Note4}. Although the pole is below the edge of the continuum, it leads to a hidden mode and no peak in $\Im \chi_H(\om+i\delta,\bq)$ (see text). The peak in $\Im \chi_H(\om+i\delta,\bq)$  at small $\omega$ corresponds to the ABG mode.}
\label{fig:chi_Ef_0.1}
\end{figure}

Our results, presented in Fig. \ref{fig:chi_Ef_0.1}, confirm this. As in Fig. \ref{fig:chi_Ef_10}(a), we plot the spectral function $\Im \chi_H(\om+i\delta,\bq)$ and overlay $\Re(z_\bq)$, where $z_\bq$ is the position of the pole of $\chi^\down_H(z,\bq)$ in the lower half-plane. We see from Fig. \ref{fig:chi_Ef_0.1} that the Higgs mode is relatively non-dispersive: for all $q$ which we study, the real part of the pole lies near $2\gap$. Although the position of the pole is only presented in Fig. \ref{fig:chi_Ef_0.1} for $q \lesssim 0.3k_F$, we find that the pole stays near $z=2\gap$ for larger values of $q$. The absence of pole positions for larger $q$ in Fig. \ref{fig:chi_Ef_0.1} is due to numerical difficulties \footnote{We obtain the position of the pole in $\chi^\down_H(z,\bq)$ by solving $\det \hat{M}^\down(z_\bq,\bq)=0$ using Newton's method. For poles sufficiently close to the real-frequency axis, we find that Newton's method does not converge. This possible failure of Newton's method for finding complex roots is well-known \cite{Epureanu1998}.}.

We see from the figure that $\Im\chi_H(\om+i\delta,\bq)$ does not display any peak. This is consistent with the hidden nature of the Higgs mode. The disappearance of the observable Higgs peak in the BEC regime, where $\mu <0$, agrees with previous theoretical results in 3D \cite{Kurkjian2019,Castin2019,Castin2020}, experimental results for 3D cold-atom systems \cite{Behrle2018}, and numerical studies in 2D \cite{Zhao2020}. As in the case of positive $\mu$, the ABG mode is visible below the two-particle continuum. Compared to Fig. \ref{fig:chi_Ef_10}, the dispersion of the ABG mode is much flatter than for $\mu > 0$. This can be understood from the mean-field dispersion of the ABG mode, $\om_\mathrm{ABG}(q) = cq$ where $c = v_F/\sqrt{2}$ \cite{Mozyrsky2019}. Since the velocity of the ABG mode is proportional to $v_F$, the dispersion of the ABG mode becomes flatter as the density decreases.

\subsection{Including the Coulomb interaction}
\label{sec:coulomb}

We now extend our analysis to account for the effects of the long-range Coulomb interaction. To do so, we return to the action. Extending Eq. (\ref{eq:action_neutral}) to include the Coulomb interaction, we have
\begin{multline}
    S[\bpsi,\psi] = \sum_k \bpsi_{k\sigma}\left(-i\om_n+
    \frac{k^2}{2m} - \mu
    \right)\psi_{k\sigma}\\
    -
    g\frac{T}{L^2}\sum_{kqp}\bpsi_{k+q/2\up}\bpsi_{-k+q/2\down}\psi_{-p+q/2\down}\psi_{p+q/2\up}\\
    +
    \frac{T}{2L^2}\sum_{pkq}\bpsi_{p+q\sigma}\bpsi_{k-q\sigma'}V_c(\bq)\psi_{k\sigma'}\psi_{p\sigma},
\end{multline}
where the Coulomb interaction in two dimensions is $V_c(\bq) = 2\pi e^2/q$. To decouple the quartic terms, we introduce two Hubbard-Stratonovich fields, $\gap$ and $\Phi$ for the particle-particle and particle-hole channels, respectively. The mean-field equations for $\Delta$ and $\Phi$, $\delta S/\delta \gap = 0$ and $\delta S /\delta \Phi = 0$, yield $\Phi = 0$ and an unchanged gap equation Eq. (\ref{eq:gap}). Similarly, the constraint of particle-number conservation yields Eq. (\ref{eq:particle_number}), as in the neutral case. Then we still have $\mu = E_F -E_0$ and $\gap = 2\sqrt{E_F E_0}$.

Of course, in reality the Coulomb repulsion does affect $\mu$ and $\gap$: it certainly weakens a system's tendency toward $s-$wave superconductivity \cite{Morel1962,Grabowski1984,Phan2022,Pimenov2022} and
may also lead to superconducting instabilities in non-$s$-wave channels \cite{Kohn1965,Maiti2013}. That $\mu$ and $\gap$ are unaffected by the repulsive Coulomb interaction in our calculation follows from the fact that we decouple the Coulomb interaction in the particle-hole channel, but not in the particle-particle channel. This is an approximation which we use simply because our goal is to analyze the effect of the Coulomb interaction on the Higgs mode.

To include fluctuations, we introduce as before, the amplitude and phase fluctuation fields, $\lambda(x)$ and $\theta(x)$. Additionally, we include fluctuations of
$\Phi(x)$ about the mean field $\Phi =0$. Expanding the action to quadratic order in $\lambda_q$, $\theta_q$, and $\Phi_q$ and integrating out $\Phi_q$ following Ref. \cite{Cea2015}, we obtain the effective action in the form

\begin{equation}
    S_\mathrm{eff} = S_0
    + \beta \gap^2 L^2
    \sum_q
    \begin{pmatrix}
    \theta_q^* & \lambda_q^*
    \end{pmatrix}
    \hat{M}(i\Om_m,\bq)
    \begin{pmatrix}
    \theta_q \\
    \lambda_q
    \end{pmatrix},
\end{equation}
where the matrix elements are now
\begin{align}
    M_{++}(z,\bq)
    &= \frac{1}{g} + \frac{1}{2}\left(\chi_{22}(z,\bq) - \frac{\chi_{23}(z,\bq)^2}{V_c^{-1}(\bq)-\chi_{33}(z,\bq)}\right)\\
    M_{--}(z,\bq)
    &= \frac{1}{g} + \frac{1}{2}\left(\chi_{11}(z,\bq) + \frac{\chi_{13}(z,\bq)^2}{V_c^{-1}(\bq)-\chi_{33}(z,\bq)}\right)\\
    \label{eq:Mmm_coulomb}
    M_{+-}(z,\bq)
    &= \frac{1}{2}\left(\chi_{12}(z,\bq)-\frac{\chi_{23}(z,\bq)\chi_{13}(z,\bq)}{V_c^{-1}(\bq)-\chi_{33}(z,\bq)} \right).
\end{align}
The susceptibilities $\chi_{11}, \chi_{22}$, and $\chi_{12}$ are the same as in Eqs. (\ref{chi_11}), (\ref{chi_22}), and (\ref{chi_12}). The new susceptibilities $\chi_{33}$, $\chi_{13}$, and $\chi_{23}$, which appear in the presence of the Coulomb interaction, are

\begin{align}
    \chi_{33}(z,\bq)
    &=
    -2 \int \frac{d^2p}{(2\pi)^2}
    \frac{E_++E_-}{2E_+E_-}\cdot
    \frac{\xi_+\xi_--E_+E_--\gap^2}{z^2-(E_++E_-)^2}\\
    \chi_{13}(z,\bq)
    &=
    2\gap \int \frac{d^2p}{(2\pi)^2}
    \frac{E_++E_-}{2E_+E_-}
    \cdot
    \frac{\xi_++\xi_-}{z^2-(E_++E_-)^2}\\
    \chi_{23}(z,\bq)
    &=
    2i\gap z \int \frac{d^2p}{(2\pi)^2}
    \frac{E_++E_-}{2E_+E_-}\cdot \frac{1}{z^2-(E_++E_-)^2}.
\end{align}

Our goal is to calculate the Higgs susceptibility $\chi_H (z,\bq) = M_{++}(z,\bq,)/\det \hat{M}(z,\bq)$ at small but finite $q$.  As in Sec. \ref{sec:neutral_analytic}, we assume that the pole is at  $z=z_\bq= 2\gap + \zeta \frac{q^2}{2m}\frac{\mu}{\gap}$ and search for a solution of $\det \hat{M}(z_\bq,\bq) = 0$. We find (see Sec. \ref{app:expansions} of the SI for details) that at arbitrary $\mu > 0$,  $M_{++}(z_\bq,\bq)$ and $M_{--}(z_\bq,\bq)$ are $O(q)$, while $M_{+-}(z_\bq,\bq) = O(q^2)$. From this, we see that the long-range Coulomb interaction effectively decouples the amplitude and phase oscillations in the long-wavelength limit, regardless of the value of $\mu/\Delta$. We then simply have

\begin{equation}
    \chi_H(z_\bq,\bq)
    \approx
    \frac{1}{M_{--}(z_\bq,\bq)}.
\end{equation}

To obtain $M_{--}(z_\bq,\bq)$ at small $\bq$, we set $\chi_{13}(z_\bq,\bq) \approx \chi_{13}(2\gap,0)$ and neglect $V_c^{-1}(\bq)$ compared to $\chi_{33} (z_\bq,\bq)$, which is $O(1/q)$. Evaluating $\chi_{33}$ in the same way as in Sec. \ref{sec:neutral_analytic} (see also Sec. \ref{app:technical_details} of the SI for a similar calculation) and using our earlier result for $\chi_{11}(z_\bq,\bq)$, we find
\begin{multline}
    M_{--}(z_\bq,\bq) \approx -iN_0 \frac{2\gap}{v_\mu q}\frac{\sqrt{\zeta}}{K(\frac{1}{\sqrt{\zeta}})}\\
    \times\left[E(\frac{1}{\sqrt{\zeta}})K(\frac{1}{\sqrt{\zeta}})+\frac{1}{16}\log(\frac{\sqrt{\gap^2+\mu^2}+\gap}{\sqrt{\gap^2+\mu^2}-\gap})^2\right].
\end{multline}

Analytically continuing $ M_{--}(z_\bq,\bq)$ into the lower half-plane for $\zeta \in (0,1)$ as we did in Sec. \ref{sec:neutral_analytic}, we obtain the Higgs susceptibility $\chi_H^\down(z_\bq,\bq)$ in the presence of the Coulomb interaction as
\begin{widetext}
\begin{multline}
    \chi^\down_H(z_\bq,\bq)
    =
    \frac{i}{N_0} \frac{2\gap}{v_\mu q}\frac{1}{\sqrt{\zeta}}\\
    \times
    \frac{K(\frac{1}{\sqrt{\zeta}})-2iK(\sqrt{1-\zeta})\sqrt{\zeta}}
    {\left[K(\frac{1}{\sqrt{\zeta}}) - 2iK(\sqrt{1-\zeta})\sqrt{\zeta}\right]
    \left[
    E(\frac{1}{\sqrt{\zeta}})+2i\left(E(\sqrt{1-\zeta^{-1}})-K(\sqrt{1-\zeta^{-1}})\right)
    \right]
    +
    \frac{1}{16}\left(\log(\frac{\sqrt{\mu^2+\gap^2}+\gap}{\sqrt{\mu^2+\gap^2}-\gap})\right)^2}.
    \label{eq:chi_mm_1}
\end{multline}
\end{widetext}
This is exactly the same equation for $\zeta$ as Eq. (\ref{eq:chi_mm}) in the absence of Coulomb interaction.
From this, we see that although the Coulomb interaction drastically modifies the character of the phase oscillations, transforming the ABG mode into the plasmon, the Higgs mode is \textit{unaffected} by the presence of the long-range Coulomb interaction.
A similar calculation shows that the Higgs mode is also unaffected by the long-range Coulomb interaction in
3D.

This result is unintuitive, since the presence of Coulomb interaction leads to a decoupling of amplitude and phase oscillations at all $\mu/\gap$. Hence, one might reasonably expect the Higgs mode to behave substantially differently in the charged system compared to a neutral superfluid. It is therefore remarkable that the location of the Higgs mode is identical in both the neutral and charged systems.

\begin{figure}
\centering
\includegraphics[width=\columnwidth]{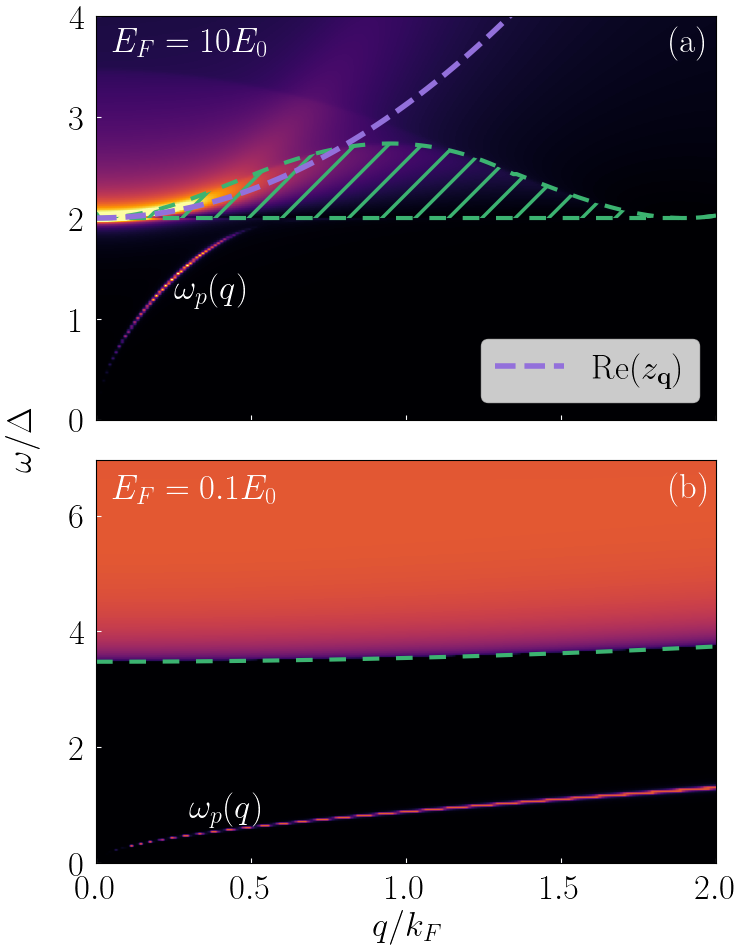}
\caption{\label{fig:coulomb} The spectral function $\Im \chi_H(\om+i\delta,\bq)$ in the charged system, for (a) $E_F = 10E_0$ and (b) $E_F = 0.1E_0$. As in Fig. \ref{fig:chi_Ef_10}, the hatched
green
region in panel (a) corresponds to frequencies between $2\gap$ and $\om_2$. Poles with $\Re(z_\bq)$ inside this hatched region lead to peaks in the spectral function. The
%DP
dashed
%%%
 purple
line-- the analytical result for the small-$q$ dispersion of the Higgs mode. In panel (b), the dashed
green
curve delineates the boundary of the two-particle continuum at $
\om_\mathrm{min} =
2\sqrt{\gap^2+(\abs{\mu}+q^2/8m)^2}$. The dispersive peak below the two-particle continuum in both panels is the plasmon mode $\om_p(q)$.
In panel (a), we scaled $\Im\chi_H(\om+i\delta,\bq)$ by a factor of 5 for $\om < 2\gap$ to enhance visibility of the plasmon mode.}
\end{figure}

We verify these analytical results by numerically calculating the spectral function $\Im\chi_H(\om+i\delta,\bq)$ in the charged system. For these calculations, we employ the dimensionless Wigner-Seitz radius, $r_s =1/(\pi n a_0^2)^{1/2}$, where $n$ is the fermionic density and $a_0 = 1/me^2$ is the Bohr radius. Recalling that $k_F = \sqrt{2\pi n}$, the Coulomb interaction in terms of $r_s$ is $V_c(q) = 2\pi e^2/q = r_s/(\sqrt{2}N_0 \Bar{q})$, where $\Bar{q} = q/k_F$. In our numerical calculations, we set $r_s=1$ when $E_F = 10E_0$. Since $r_s \sim 1/\sqrt{E_F}$, $r_s$ at any other Fermi energy can be obtained through $r_s(E_F) = \sqrt{10E_0/E_F}$.

In Fig. \ref{fig:coulomb}, we plot $\Im \chi_H(\om+i\delta,\bq)$ in the charged system, using $E_F = 10E_0$ in panel (a) and $E_F = 0.1E_0$ in panel (b). Compared with Fig. \ref{fig:chi_Ef_10}(a), the only significant difference in the spectral function is in the ABG mode. The ABG mode, which disperses linearly with $q$ in the neutral system, transforms into the plasmon in the charged system, which disperses as $\sqrt{q}$ in 2D. More drastically, we find that the inclusion of the Coulomb interaction leads to dramatic depletion of the spectral weight of the gapless mode, especially in the high-density case of Fig. \ref{fig:coulomb}(a).
To make the plasmon mode visible in Fig. \ref{fig:coulomb}(a), we multiplied $\Im\chi_H(\om+i\delta,\bq)$ by a factor of 5 for $\om < 2\gap$.
This highlights the significant decoupling of amplitude and phase fluctuations in the charged system, and is fully consistent with our analytical treatment. Moreover, our numerical results show that the decoupling is not restricted to only small $q$, but persists to substantially larger $q \geq k_F$.

In the case of Fig. \ref{fig:coulomb}(b) where we are in the BEC regime, we also find that the Higgs mode is not affected by the Coulomb interaction. In particular, just as in the neutral superfluid (c.f. Fig. \ref{fig:chi_Ef_0.1}), there is no peak in $\Im\chi_H(\om+i\delta,\bq)$ which can be attributed to the Higgs mode. This agrees with the results of Ref. \cite{Cea2015}. Instead, we only have a plasmon peak below the two-particle continuum.

\section*{\uppercase{Discussion}}
\label{sec:discussion}
In this work, we obtained the dispersion, damping rate, and residue of the Higgs mode in two dimensions across the BCS-BEC crossover and analyzed under which conditions this mode gives rise to a peak in the imaginary part of the Higgs susceptibility, $\Im \chi_H(\om+i\delta,\bq)$, a quantity which is observable using spectroscopic probes.

To detect the Higgs mode, we calculated the Higgs susceptibility $\chi_H(z,\bq)$ in the upper half-plane of complex $z$ and obtained its analytic continuation $\chi_H^\down(z,\bq)$ into the lower half-plane. We found that $\chi_H^\down(z,\bq)$ has a pole (the Higgs mode), whose location in the lower half-plane is $z_\bq = 2\gap + (0.5-i\beta)\frac{q^2}{2m}\frac{\mu}{\gap}$ for $\mu > 0$ and small $\bq$. Here, the damping parameter $\beta$ is given by $\beta = 0.4308$ at $\mu \gg \gap$ and diverges as $\frac{e}{16}\sqrt{\frac{2\gap}{\mu}}$ for $\mu \ll \gap$. Additionally, we calculated the residue $Z_\bq$ of the pole, finding that $Z_\bq$ scales linearly with $q$, and goes to zero at $\mu = 0$ as $(\mu/2\gap)^{1/4}$. We found that for small $q$, the Higgs mode gives rise to a peak in the observable $\Im\chi_H (\omega +i\delta, \bq)$ for any positive value of the dressed chemical potential $\mu$. We then numerically obtained the position of the pole at larger $q$, finding that
the Higgs mode does not give rise to a peak in $\Im\chi_H (\omega +i\delta, \bq)$ once $q$ crosses some threshold value. For negative $\mu$, we found that the Higgs mode is hidden below a branch cut, and does not lead to any peak in $\Im\chi_H(\om+i\delta,\bq)$. Lastly, we included the effect of the long-range Coulomb interaction and demonstrated that its inclusion does not affect the Higgs mode, despite the fact that it decouples the phase (density) and amplitude channels.

A final note: in this work we only decoupled our attractive Hubbard interaction in the particle-particle channel, neglecting its effect on the particle-hole channel. This is valid in the high-density or weak-coupling limits, where particle-hole symmetry holds. Away from these limits, renormalization of $\chi_H(z,\bq)$ in the particle-hole channel from the Hubbard interaction, similar to our treatment of the Coulomb interaction in Sec. \ref{sec:coulomb}, is likely necessary to obtain the correct dispersion of the Higgs mode~\cite{lara_1}.

%DP
In summary, our work adds to a growing corpus of studies which analyze the coupling between collective modes and a continuum of single-particle excitations \cite{Klimin2019,Klimin2019leggett,Lumbeeck2020,Kurkjian2020,Klimin2021,Repplinger2022,Klimin2022}. The generality of the techniques employed here suggests that analytical continuation may be helpful in the study of other physical problems, such as that of plasmon decay inside the particle-hole continuum of strange metals \cite{Wang2022}.

\section*{\uppercase{Methods}}
All calculations not performed in the main text are detailed in the SI.

\paragraph*{\bf{Acknowledgment}}
We acknowledge useful conversations with L. Benfatto, D. Chowdhury, and P. Littlewood. This work was supported by the U.S. Department of Energy, Office of Science, Basic Energy Sciences, under Award No. DE-SC0014402.

\paragraph*{\bf{Data availability}}
Data will be kept in a UMN database, and is available upon request.

\paragraph*{\bf{Author contributions}}
A.V.C. designed the project. D.P. performed the calculations with input from A.V.C. The authors discussed the results, their relation to experiments, and wrote the manuscript together.

\paragraph*{\bf{Competing interests}} The authors declare no competing interests.

\pagebreak

\onecolumngrid

\section*{\uppercase{Supplementary Information}}

\appendix

\renewcommand{\thesubsection}{\Alph{subsection}}

\subsection{Evaluation of $M_{--}(z,\bq)$ at $\mu > 0$ and small $q$}
\label{app:technical_details}

In this section, we provide the details on the calculation of the matrix elements $M_{--}(z,\bq)$ at small $q$ and $\mu > 0$. We begin by calculating $M_{--}(z,\bq)$ given by

\begin{equation}
    M_{--}(z,\bq)
    =
    \frac{1}{4}\int \frac{d^2p}{(2\pi)^2}
    \frac{E_++E_-}{E_+E_-}\cdot
    \frac{z^2-4\gap^2-(\xi_+-\xi_-)^2}{z^2-(E_++E_-)^2}.
\end{equation}

For $\mu > 0$ and small $\bq$, we expect the majority of the weight in this integral to come from momenta near $p = p_\mu = \sqrt{2m\mu}$. As such, we expand the quantities entering the integrand to quadratic order in $\bq$ and $\delta p \equiv p - p_\mu$. Defining $v_\mu = p_\mu/m$, we have $(\xi_+-\xi_-)^2 \approx v_\mu^2 q^2\cos^2\theta$ in the numerator, and $(E_++E_-)^2 \approx 4\gap^2 + v_\mu^2q^2\cos^2\theta+4v_\mu^2\delta p^2$ in the denominator. The rest of the integrand is nonsingular, and can be evaluated at $\bq = \delta p = 0$. With this, we use $d^2p \approx p_\mu d\delta p d\theta$ and obtain

\begin{equation}
    M_{--}(z,\bq)
    \approx
    \frac{p_\mu}{8\pi^2\gap}\int_0^{2\pi} d\theta \int_{-\infty}^\infty d\delta p
    \frac{z^2-4\gap^2-v_\mu^2 q^2\cos^2\theta}{z^2-4\gap^2 -v_\mu^2q^2\cos^2\theta-4v_\mu^2\delta p^2}.
\end{equation}

Before further evaluation of this expression, we note that the denominator in the integrand is highly singular, since the region of interest corresponds to small $q$, small $\delta p$, and $z \approx 2\gap$. As such, one must be careful when simplifying this expression. In Ref. \cite{Littlewood1982}, the authors evaluated $M_{--}(z,\bq)$ (in three dimensions) by setting $q=0$ in the denominator. This led the authors to a dispersion of the Higgs mode which disagrees with the results of Andrianov and Popov, who in contrast did not set $q=0$ in the denominator.

With that said, we now continue our evaluation of $M_{--}(z,\bq)$ by performing the integral over $\delta p$, finding

\begin{equation}
    M_{--}(z,\bq)
    \approx
    -i \frac{N_0}{8\gap}\int_0^{2\pi} d\theta
    \sqrt{z^2-4\gap^2-v_\mu^2 q^2\cos^2\theta},
\end{equation}

where $N_0 = m/2\pi$ is the density of states per spin in two dimensions. Defining the dimensionless parameter $\zeta = (z^2-4\gap^2)/v_\mu^2q^2$, $M_{--}(z,\bq)$ becomes

\begin{equation}
    M_{--}(z,\bq)
    =
    -i N_0 \frac{v_\mu q}{2\gap}\sqrt{\zeta}E(\frac{1}{\sqrt{\zeta}}),
\end{equation}

where $E(z)$ is the complete elliptic integral of the second kind. We note that this definition of $\zeta$ agrees with the definition of $\zeta$ in Sec. \ref{sec:neutral_analytic} of the main text, where $z = 2\gap + \zeta \frac{q^2}{2m}\frac{\mu}{\gap}$, since we are working at small $\bq$.

\subsection{Lifshitz transitions in the matrix elements $M_{\sigma\sigma'}(\om,\bq)$}
\label{app:analytic_continuation_discussion}
\begin{figure*}
\includegraphics[width=0.95\textwidth]{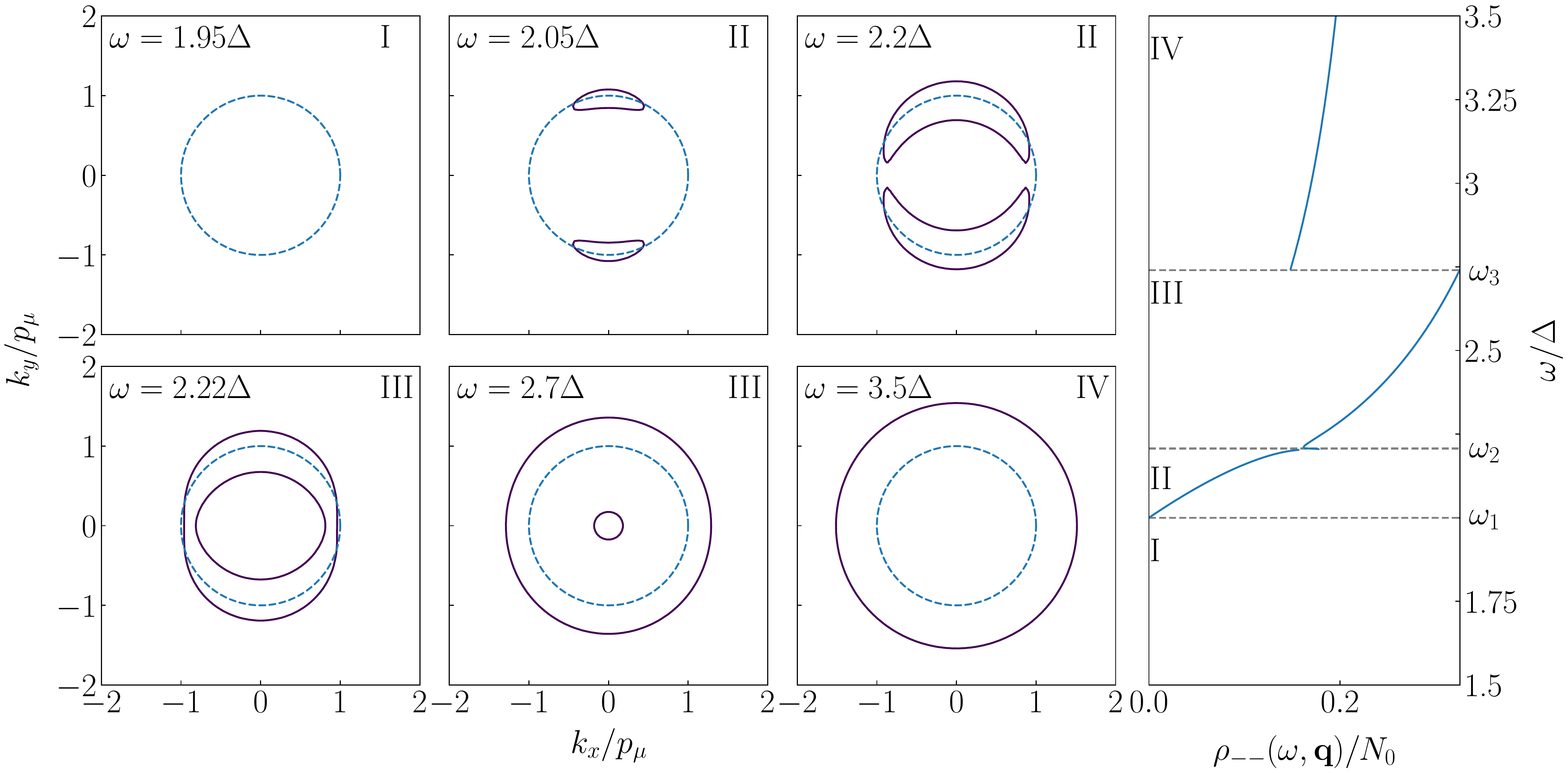}
\caption{\label{fig:topology} We plot for different values of $\om$ the momenta $\bk$ which satisfy $\om = E(\bk+\bq/2) + E(\bk-\bq/2)$. In the top right of each panel, we label the region of analyticity that each $\om$ corresponds to. A dashed blue curve where $k = p_\mu \equiv \sqrt{2m\mu}$ has been added to each panel for reference. In the far-right panel, we have plotted $\rho_{--}(\om,\bq)$ as a function of $\om$, highlighting the boundaries between the various regions with dashed lines. Here, we have taken $\mu = \gap$ and $\bq = 0.5p_\mu \hat{x}$.}
\end{figure*}

In this section, we discuss the topological origin of the kinks and discontinuities in the spectral densities $\rho_{\sigma\sigma'}(\om,\bq)$ and matrix elements $M_{\sigma\sigma'}(\om,\bq)$. In the far-right panel of Fig. \ref{fig:topology}, we plot the behavior of $\rho_{--}(\om,\bq)$ as a function of $\om$, taking for concreteness $\mu = \gap$ and $q = 0.5 p_\mu$, where we introduce for convenience $p_\mu \equiv \sqrt{2m \mu}$. As is evident from the figure, $\rho_{--}(\om,\bq)$ is not smooth as a function of $\om$ (these kinks and discontinuities were also briefly discussed in Sec. \ref{sec:ac_overview} of the main text.)

For $\om < \om_1 = 2\gap$, $\rho_{--}(\om,\bq)$ is zero. As $\om$ increases, a ``kink'' appears $\rho_{--}(\om,\bq)$ at $\om_2$, as well as a discontinuous jump in $\rho_{--}(\om,\bq)$ at $\om_3$ \footnote{Unlike in two dimensions, $\rho_{--}(\om,\bq)$ is continuous across $\om_3$ in three dimensions. This behavior is analogous to the density of states: in three dimensions, the density of states continuously goes to zero as $\sqrt{\varepsilon}$ at we approach the bottom of the band; in two dimensions, the density of states is constant as we approach the bottom of the band, jumping to zero when we go below $\varepsilon=0$.}. The frequencies $\om_1$, $\om_2$, and $\om_3$ mark transition points where the analytic behavior of $\rho_{--}(\om,\bq)$ changes. To highlight the different analytic behaviors in $\rho_{--}(\om,\bq)$, we introduce four regions of $\om$: Region I, where $\om < 2\gap$; Region II, where $\om \in (2\gap,\om_2(\bq))$; Region III, where $\om \in (\om_2(\bq), \om_3(\bq))$; and Region IV, where $\om > \om_3(\bq)$.

These transitions in the behavior of $\rho_{--}(\om,\bq)$ between the various regions is in fact topological in nature. To see this, consider the expression for $\rho_{--}(\om,\bq)$

\begin{equation}
    \rho_{--}(\om,\bq) = \int \frac{d^2k}{(2\pi)^2}
    \frac{\xi_+\xi_- + E_+E_--\gap^2}{4E_+E_-}\delta(\om-E_+-E_-),
    \label{eq:rho_mm}
\end{equation}
which can be obtained from Eq. (\ref{eq:discontinuity_cut}), Eq. (\ref{eq:Mmm}) and Eq. (\ref{chi_11}) of the main text. From this expression, the only momenta $\bk$ which contribute to $\rho_{--}(\om,\bq)$ must satisfy $\om = E_++E_-$, or more explicitly, $\om = \sqrt{\xi(\bk+\bq/2)^2+\gap^2} + \sqrt{\xi(\bk-\bq/2)^2+\gap^2}$. In Fig. \ref{fig:topology}, we plot in purple the momenta $\bk$ which contribute to $\rho_{--}(\om,\bq)$ for various values of $\om$. We also add for reference a dashed-blue circle where $k = p_\mu$.

For $\om < 2\gap$, there are no momenta which satisfy $\om = E_++E_-$, since the minimum value $E(\bk)$ can take is $\gap$ (for $\mu > 0$.) As such, $\rho_{--}(\om,\bq) = 0$ for $\om < 2\gap$ in Region I. As $\om$ increases past $2\gap$, the set of momenta which satisfy $\om = E_++E_-$ becomes nonempty, and we enter Region II. We see from Fig. \ref{fig:topology} that the momenta satisfying $\om = E_++E_-$ in Region II begin as ``bubbles'' localized around $\bk = \pm p_\mu \hat{y}$. These bubbles grow with increasing $\om$, and eventually reconfigure into two concentric closed curves as $\om$ crosses $\om_2$ and we enter Region III. In Region III, increasing $\om$ leads to a gradual increase in the size of the outer curve, and a shrinking of the inner curve. This inner curve disappears as $\om$ crosses $\om_3$, and we enter Region IV. In Region IV, increasing $\om$ simply leads to a increase in the size of the remaining curve.

In the case where $\mu < 0$, one finds that Regions II and III disappear. Instead, $\rho_{--}(\om,\bq) = 0$ for $\om < \om_3$ (same as $\om_\mathrm{min}$ used in the main text), and discontinuously jumps from 0 to a finite value as $\om$ crosses $\om_3$ into Region IV. Topologically, this jump in $\rho_{--}(\om,\bq)$ arises from the set of momenta satisfying $\om=E_++E_-$ becoming nonempty, forming a closed curve enclosing $\bk=0$. Increasing $\om$ leads to an increase in the size of this curve.

\subsection{Analytical continuation of the matrix elements for arbitrary $\mu$, $\om$, and $\bq$}
\label{app:analytic_continuation_full}
In this section, we provide details for the calculation of the spectral densities $\rho_{\sigma\sigma'}(\om,\bq)$, and discuss their analytical continuation away from the real-frequency axis. In doing so, we follow Refs. \cite{Castin2019,Castin2020} closely, making minimal changes to work in two dimensions. In general, we have

\begin{equation}
    \rho_{\sigma \sigma'}(\om,\bq) = \int \frac{d^2k}{(2\pi)^2} m_{\sigma\sigma'}(\bk,\bq)\delta(\om - E_+-E_-).
\label{eq:spectral_def}
\end{equation}

where the form of $m_{\sigma\sigma'}(\bk,\bq)$ depends on which matrix element is under consideration. For example, from Eq. (\ref{eq:rho_mm}), we have $m_{--}(\bk,\bq) = (\xi_+\xi_- + E_+E_--\gap^2)(4E_+E_-)$ in the neutral system. For convenience, we rescale the momenta and energies to be in terms of $k_F$ and $E_F$, respectively. This corresponds to working with the dimensionless variables $k = k_F \Bar{k}$, $\xi = E_F \Bar{\xi}$, etc. The spectral density $\rho_{\sigma \sigma'}(\om,\bq)$ is then given by

\begin{equation}
    \rho_{\sigma \sigma'}(\om,\bq) = \frac{k_F^2}{E_F}\int \frac{d^2\Bar{k}}{(2\pi)^2} m_{\sigma\sigma'}(\Bar{\bk},\Bar{\bq})\delta(\Bar{\om} - \Bar{E}_+-\Bar{E}_-).
\end{equation}

Using $E_F = k_F^2/2m$ and $N_0 = m/2\pi$, we rewrite the prefactor of the integral as $k_F^2/E_F = 4\pi N_0$. To proceed, we note that the $\rho_{\sigma \sigma'}(\om,\bq)$ is invariant under $\bq \rightarrow -\bq$, due to the inversion symmetry present in the system. Mathematically, this arises from the symmetry of the integrand, which is even in $\cos\theta$ for all $\sigma$ and $\sigma'$, in both the neutral and charged system. As such, in all angular integrals, we take $\int_0^{2\pi}d\theta\rightarrow 4\int_0^{\pi/2}d\theta$, where $\cos\theta \in (0,1)$. To further simplify $\rho_{\sigma \sigma'}(\om,\bq)$, we rewrite the delta function in a more suitable form. We follow Refs. \cite{Castin2019,Castin2020}, and introduce the quantities $\Bar{\xi}_q$, $R$, and $\theta_r$, defined as

\begin{align}
    \Bar{\xi}_q &= \Bar{k}^2-\Bar{\mu}+\Bar{q}^2/4\\
    R &= \frac{(\Bar{\om}^2-4\Bar{\gap}^2)/4-\Bar{\xi}_q^2}{\Bar{\om}^2/4-\Bar{\xi}_q^2}\\
    \cos(\theta_r(\bark)) &= \frac{\Bar{\om}}{2\Bar{k}\Bar{q}}\sqrt{R}.
\end{align}

Although these three quantities are clearly functions of $\bark$ by definition, we write $\theta_r = \theta_r(\bark)$ to emphasize its functional dependence. With these definitions, one can show that (see Ref. \cite{Castin2019} for details)

\begin{equation}
    \delta(\Bar{\om} - \Bar{E}_+-\Bar{E}_-)
    =
    \frac{1}{2\Bar{k}\Bar{q}\sqrt{R}}
    \frac{\Bar{\om}^2/4-\Bar{\xi}_q^2R}{\Bar{\om}^2/4-\Bar{\xi}_q^2}
    \delta(\cos\theta-\cos\theta_r(\bark)).
\end{equation}

From the delta function on the right-hand side, we see that $\cos\theta_r(\bark)$ must be a real number between $0$ and $1$ (recall that by exploiting the inversion symmetry of the system, we only need to consider values of $\theta$ between 0 to $\pi/2$, where $0<\cos\theta<1$.) Constraining $\cos\theta_r(\bark)$ to be real is equivalent to only considering $\bxi_q \in (-\frac{\sqrt{\bom^2-4\bgap^2}}{2},\frac{\sqrt{\bom^2-4\bgap^2}}{2})$, while the constraint that $\cos\theta_r(\bark) \in (0,1)$ is enforced by appending $\Theta(1-\cos\theta_r(\bark))$ to the right-hand side. Using this, we perform the integral over $\theta$, obtaining

\begin{equation}
    \rho_{\sigma\sigma'}(\om,\bq)
    =
    \frac{4N_0}{\pi}\int_0^\infty \Bar{k} d\Bar{k}
    m_{\sigma\sigma'}(\bk,\bq)
    \frac{1}{2\Bar{k}\Bar{q}\sqrt{R}}
    \frac{\Bar{\om}^2/4-\Bar{\xi}_q^2R}{\Bar{\om}^2/4-\Bar{\xi}_q^2}
    \frac{1}{\sin\theta_r(\bark)}\Theta(1-\cos\theta_r(\bark))\Theta(\bom^2/4-\bgap^2-\bxi_q^2).
    \label{eq:rho_kindof_simplified}
\end{equation}

To simplify this integral, we now write out $m_{\sigma\sigma'}(\bk,\bq)$. When the delta function is satisfied, one can show that $\bE_\pm = \bom/2 \pm \bxi_q \sqrt{R}$ and $\bxi_\pm = \bxi_q \pm \bom \sqrt{R}/2$. As such, we find (for the neutral superfluid)

\begin{align}
    m_{++}(\bk,\bq)
    &=
    \frac{\bgap^2\bom^2/4}{(\bom^2/4-\bxi_q^2)(\bom^2/4-\bxi_q^2R)}
    \\
    m_{--}(\bk,\bq)
    &=
    \frac{\bgap^2\bxi_q^2}{(\bom^2/4-\bxi_q^2)(\bom^2/4-\bxi_q^2R)}
    \\
    m_{+-}(\bk,\bq)
    &=
    -i\frac{\bgap^2 \bom \bxi_q}{2(\bom^2/4-\bxi_q^2)(\bom^2/4-\bxi_q^2 R)}.
\end{align}

For concreteness, we now specialize to $\rho_{+-}(\om,\bq)$; the other spectral densities are obtained in a similar manner. Inserting the above expression for $m_{+-}(\bk,\bq)$, we find after some work that

\begin{align}
    \rho_{+-}(\om,\bq)
    &=
    -i\frac{\bgap^2 \bom N_0}{4\pi \barq}\int_{-\sqrt{\bom^2/4-\bgap^2}}^{\sqrt{\bom^2/4-\bgap^2}} d\bxi_q
    \frac{ \bxi_q}{(\bom^2/4-\bxi_q^2)^{3/2}}
     \frac{1}{( \bom^2/4-\bgap^2-\bxi_q^2)^{1/2}}\frac{1}{(\bxi_q+(\bmu-\barq^2/4)-\frac{\bom^2}{4\barq^2}R)^{1/2}}\Theta(1-\cos\theta_r).
\end{align}

Note that in the above expression, we have switched to integrating over $\bxi_q$. We now perform a further change of variables, $\bxi_q = \sqrt{\bom^2/4-\bgap^2}s$, after which, the integral becomes

\begin{multline}
    \rho_{+-}(\om,\bq)
    =
    -i\frac{\bgap^2 \bom \sqrt{\bom^2/4-\bgap^2}N_0}{4\pi}\int_{-1}^1 ds
    \frac{ \Theta(1-\cos\theta_r)}{\bom^2/4-(\bom^2/4-\bgap^2)s^2}\frac{s}{( 1-s^2)^{1/2}}\\
     \times\frac{1}{\left[\barq^2\left(\bom^2/4-(\bom^2/4-\bgap^2)s^2\right)\left(s\sqrt{\bom^2/4-\bgap^2}+\bmu-\barq^2/4\right)-\bom^2/4(\bom^2/4-\bgap^2)(1-s^2)\right]^{1/2}}.
     \label{eq:rho_pm_full}
\end{multline}

The quantity under the right-most square root corresponds to the $\sin\theta_r = \sqrt{1-\cos^2\theta_r}$ in the denominator of Eq. (\ref{eq:rho_kindof_simplified}). As such, the step function $\Theta(1-\cos\theta_r)$ merely ensures that the quantity under this right-most square root is positive. Therefore, this constraint is equivalent to the following:

\begin{equation}
    \sqrt{\bom^2/4-\bgap^2}s
    +\bmu-\barq^2/4
    >\frac{\bom^2}{4\barq^2}\frac{1-s^2}{\frac{\bom^2/4}{\bom^2/4-\bgap^2}-s^2}
\end{equation}

The bounds of the integral for $\rho_{+-}$ are then obtained by setting the left-hand and right-hand side equal. This leads to a cubic equation, the solutions of which are given by

\begin{equation}
    s_n = \frac{Y}{12\barq^2\sqrt{\bom^2/4-\bgap^2}}-\frac{1}{6\barq^2}\frac{\left(Y^2+12\bom^2\barq^4\right)^{1/2}}{\sqrt{\bom^2/4-\bgap^2}}\cos(\theta_0 + \frac{2 \pi n}{3}-\frac{\pi}{3}),
    \label{eq:roots}
\end{equation}

where $n = 0, 1, 2$ and we have introduced the quantities $Y = \bom^2+\barq^2(\barq^2-4\bmu)$, and $\theta_0$, given by

\begin{equation}
    \theta_0
    = \frac{1}{3}\arccos(\frac{Y(Y^2-36\bom^2\barq^4)+216\bom^2\barq^4\bgap^2}{\left(Y^2+12\bom^2\barq^4\right)^{3/2}}).
\end{equation}

For sufficiently small $q$ (see Ref. \cite{Castin2019} for details) and $\bmu > 0$, we have the following regions:

\begin{enumerate}[I.]
    \item $\om \in (0,2\gap)$: In this case, the spectral density is zero. This is easily seen from the original definition of the spectral density, Eq. (\ref{eq:spectral_def}), which contains $\delta(\om-E_+-E_-)$. Since the minimum value of $E_++E_-$ is $2\gap$ (or $2\sqrt{\gap^2+(\mu-q^2/8m)^2}$ for $\mu < 0$), the delta function is never satisfied if $\om < 2\gap$.
    \item $\om \in (2\gap,\omega_2(\bq))$. Of the three roots, $s_0$, $s_1$, $s_2$, only $s_2$ is real. This real root $s_2$ exists at $s_2>1$. In this case, the integral over $s$ runs over $[-1,1]$.
    \item $\om \in (\omega_2(\bq),\omega_3(\bq))$. In this case, there are two roots $s_0$ and $s_1$ in $[-1,1]$, and the integral is over $[-1,s_0]\cup [s_1,1]$.
    \item $\om > \omega_3(\bq)$. In this case, $s_0$ dips below $-1$, so the integral runs over $[s_1,1]$.
\end{enumerate}

At small $\bq$ and $\mu > 0$, one can show that $\om_2(\bq) = 2\bgap + \barq^2\frac{\bmu}{\bgap}$ (or in dimension-full variables, $\om_2(\bq) = 2\gap + \frac{q^2}{2m}\frac{\mu}{\gap}$ \cite{Castin2020}.) This result is used in Sec. \ref{sec:neutral_analytic} of the main text when we analytically continue $\chi_H(z,\bq)$ at small $\bq$. For any value of $q$ and $\mu$, we have (in dimension-full variables) $\om_3(\bq) = 2\sqrt{\gap^2+(\mu-q^2/8m)^2}$.

If $\mu < 0$ or $q$ is sufficiently large ($\barq^2/
4 > \bmu$), then the discriminant of the cubic changes sign and there is only one real root. This leads to the disappearance of Regions II and III. Instead, we have only Region I ($\om < \om_3(\bq)$), and Region IV ($\omega > \omega_3(\bq)$.) For a more thorough discussion of the roots $s_i$ and the regions of analyticity, we refer the reader to Ref. \cite{Castin2019}.

We now specialize to Region II, which is the region considered for most of this work. In this case, we have $\mu > 0$, $\barq^2/4 < \bmu$, $\om \in (2\gap,\omega_2(\bq))$, and we can drop the step function from the integral in Eq. (\ref{eq:rho_pm_full}). We find

\begin{multline}
    \rho^\mathrm{II}_{+-}(\om,\bq)
    =
    -i\frac{\bgap^2 \bom \sqrt{\bom^2/4-\bgap^2}N_0}{4\pi}\int_{-1}^1 ds
    \frac{1}{\bom^2/4-(\bom^2/4-\bgap^2)s^2}\frac{s}{( 1-s^2)^{1/2}}\\
     \times\frac{1}{\left[\barq^2\left(\bom^2/4-(\bom^2/4-\bgap^2)s^2\right)\left(s\sqrt{\bom^2/4-\bgap^2}+\bmu-\barq^2/4\right)-\bom^2/4(\bom^2/4-\bgap^2)(1-s^2)\right]^{1/2}}.
\end{multline}

This expression corresponds to a hyper-elliptic integral, which in general has no closed form expression. We therefore evaluate this integral numerically. To remove the singularities at $s=\pm 1$, we switch variables using $s=\sin\alpha$ and rewrite $\rho^\mathrm{II}{+-}(\om,\bq)$ as

\begin{multline}
    \rho^\mathrm{II}_{+-}(\om,\bq)
    =
    -i\frac{\bgap^2 \bom \sqrt{\bom^2/4-\bgap^2}N_0}{4\pi}\int_{-\pi/2}^{\pi/2} d\alpha
    \frac{\sin\alpha}{\bom^2/4-(\bom^2/4-\bgap^2)\sin^2\alpha}\\
     \times\frac{1}{\left[\barq^2\left(\bom^2/4-(\bom^2/4-\bgap^2)\sin^2\alpha\right)\left(\sin\alpha\sqrt{\bom^2/4-\bgap^2}+\bmu-\barq^2/4\right)-\bom^2/4(\bom^2/4-\bgap^2)\cos^2\alpha\right]^{1/2}}.
\end{multline}

The analytic continuation of this expression to complex frequencies is trivially obtained by taking $\om \rightarrow z$. Arguing similarly, $\rho^\mathrm{II}_{++}(z,\bq)$ and $\rho^\mathrm{II}_{--}(z,\bq)$ in the neutral system are given by

\begin{align}
\rho^\mathrm{II}_{++}(z,\bq) &=
\begin{multlined}[t]
    \frac{\bgap^2 z^2 N_0}{8\pi}\int_{-\pi/2}^{\pi/2} d\alpha
    \frac{1}{z^2/4-(z^2/4-\bgap^2)\sin^2\alpha}\\
     \times\frac{\Theta(1-\cos\theta_r)}{\left[\barq^2\left(z^2/4-(z^2/4-\bgap^2)\sin^2\alpha\right)\left(\sin\alpha\sqrt{z^2/4-\bgap^2}+\bmu-\barq^2/4\right)-z^2/4(z^2/4-\bgap^2)\cos^2\alpha\right]^{1/2}}
\end{multlined}\\
\rho^\mathrm{II}_{--}(z,\bq) &=
\begin{multlined}[t]
    \frac{\bgap^2 (z^2/4-\bgap^2)N_0}{2\pi}\int_{-\pi/2}^{\pi/2} d\alpha
    \frac{\sin^2\alpha}{z^2/4-(z^2/4-\bgap^2)\sin^2\alpha}\\
     \times\frac{\Theta(1-\cos\theta_r)}{\left[\barq^2\left(z^2/4-(z^2/4-\bgap^2)\sin^2\alpha\right)\left(\sin\alpha\sqrt{z^2/4-\bgap^2}+\bmu-\barq^2/4\right)-z^2/4(z^2/4-\bgap^2)\cos^2\alpha\right]^{1/2}}.
\end{multlined}
\end{align}

In the case where $\mu < 0$, $\rho_{\sigma\sigma'}(\om,\bq)$ is only nonzero in Region IV, where the integral over $s$ runs from $s_1$ to $1$. The above expressions for $\rho^\mathrm{II}_{\sigma\sigma'}(z,\bq)$ can be modified to obtain the spectral densities in Region IV by replacing the lower bound of the integral from $\alpha=-\pi/2$ to $\alpha = \arcsin(s_1)$, where $s_1$ is obtained from Eq. (\ref{eq:roots}). As in Region II, we can analytically continue this expression away from the real-frequency axis by taking $\om\rightarrow z$, not only in the integrand, but also in the lower bound of the integral, $\arcsin(s_1)$. This procedure allows one to analytically continue the susceptibilities $\chi_{ij}(z,\bq)$ through any desired region.

\subsection{Analytic continuation of $E(1/\sqrt{\zeta})$ and $K(1/\sqrt{\zeta})$}
\label{app:analytic_continuation_EK}
In this section, we provide details on the analytic continuation of $E(1/\sqrt{\zeta})$ and $K(1/\sqrt{\zeta})$ through the interval of the real-$\zeta$ axis where $\Re(\zeta) \in (0,1)$. To this end, we first calculate the discontinuities in these functions across this interval. Denoting $x_\pm = x \pm i \delta$ where $x \in (0,1)$, we have

\begin{align}
    K(\frac{1}{\sqrt{x \pm i \delta}})
    &=
    \int_0^{\pi/2}\frac{d\theta}{\sqrt{1- \frac{\sin^2\theta}{x \pm i \delta}}}.
\end{align}

Defining $\theta_0 = \arcsin(\sqrt{x})$, we can split up the integral into

\begin{align}
    K(\frac{1}{\sqrt{x \pm i \delta}})
    &=
    \int_0^{\theta_0} \frac{d\theta}{\sqrt{1- \frac{\sin^2\theta}{x}}}
    \mp i
    \int_{\theta_0}^{\pi/2} \frac{d\theta}{\sqrt{\frac{\sin^2\theta}{x}-1}},
\end{align}

which implies that the discontinuity across the interval of the real axis where $x \in (0,1)$ is given by

\begin{align}
    K(\frac{1}{\sqrt{x + i \delta}})-K(\frac{1}{\sqrt{x - i \delta}})
    &=
    -2i
    \int_{\theta_0}^{\pi/2} \frac{d\theta}{\sqrt{\frac{\sin^2\theta}{x}-1}}\\
    &=
    -2i\sqrt{x}K(\sqrt{1-x}).
\end{align}

To obtain the second expression, we made the substitution $\cos\theta = \sqrt{1-x}\sin\theta'$, rewriting the result in terms of elliptic integrals. From this, it follows that the analytic continuation of $K(1/\sqrt{\zeta})$ into the lower half-plane through the interval of the real-$\zeta$ axis where $\zeta \in (0,1)$ is obtained by taking $K(1/\sqrt{\zeta}) \rightarrow K(1/\sqrt{\zeta})-2i\sqrt{\zeta}K(\sqrt{1-\zeta})$, as stated in the main text. Working similarly for $E(1/\sqrt{\zeta})$, we find

\begin{align}
    E(\frac{1}{\sqrt{x + i \delta}})-E(\frac{1}{\sqrt{x - i \delta}})
    &=
    \frac{2i}{\sqrt{x}}\int_{\theta_0}^{\pi/2}d\theta \sqrt{\sin^2\theta-x}\\
    &=
    \frac{2i(1-x)}{\sqrt{x}}\int_{0}^{\pi/2}d\theta \frac{\cos^2\theta}{\sqrt{1-(1-x)\sin^2\theta}}.
\end{align}

This integral can be rewritten in terms of elliptic integrals, and one finds that

\begin{equation}
    E(\frac{1}{\sqrt{x + i \delta}})-E(\frac{1}{\sqrt{x - i \delta}})
    =
    2i(E(\sqrt{1-\zeta^{-1}})-K(\sqrt{1-\zeta^{-1}})).
\end{equation}

Adding this to $E(\frac{1}{\sqrt{\zeta}})$ when $\zeta$ is in the lower half-plane, we obtain the analytic continuation of $E(\frac{1}{\sqrt{\zeta}})$ stated in the main text.

\subsection{A Proof of the Reflection Symmetry of Eq. \ref{eq:zeta}}
\label{app:reflection_proof}

In this section, we prove the lemma stated in Sec. \ref{sec:arbitrary_mu} of the main text. That is, given a solution $\zeta$ of the equation

\begin{equation}
    \left[K(\frac{1}{\sqrt{\zeta}}) - 2iK(\sqrt{1-\zeta})\sqrt{\zeta}\right]
    \left[
    E(\frac{1}{\sqrt{\zeta}})+2i\left(E(\sqrt{1-\zeta^{-1}})-K(\sqrt{1-\zeta^{-1}})\right)
    \right]
    +
    \frac{1}{16}\left(\log(\frac{\sqrt{\mu^2+\gap^2}+\gap}{\sqrt{\mu^2+\gap^2}-\gap})\right)^2 = 0,
    \label{eq:zeta_app}
\end{equation}

then the reflection of $\zeta$ across the $\Re(\zeta) = 0.5$ line (i.e. $1-\zeta^*$) is also a solution. To rephrase this lemma, define $I(\zeta)$ to be the product on the left-hand side of the above equation, and $C = \frac{1}{16}\left(\log(\frac{\sqrt{\mu^2+\gap^2}+\gap}{\sqrt{\mu^2+\gap^2}-\gap})\right)^2 $. The above equation is then equivalent to $I(\zeta) + C= 0.$ The lemma then states that $I(1-\zeta^*)+ C = 0$ also holds. To prove this, we note that if $I(\zeta) + C = 0$, then its complex conjugate $(I(\zeta))^* + C = 0$ also holds. The lemma then follows if $(I(\zeta))^* = I(1-\zeta^*)$. To prove this, we simply calculate $I(1-\zeta^*)$. To illustrate the general procedure, consider the first term appearing in $I(\zeta)$, i.e. $K(1/\sqrt{\zeta})$. Taking $\zeta \rightarrow 1-\zeta^*$, $K(1/\sqrt{\zeta})$ becomes

\begin{align}
    K(\frac{1}{\sqrt{1-\zeta^*}})
    &=
    \int_0^{\pi/2}\frac{d\theta}{\sqrt{1-\frac{\sin^2\theta}{1-\zeta^*}}}\\
    &=
    \int_0^{\pi/2}d\theta \sqrt{\frac{1-\zeta^*}{\cos^2\theta-\zeta^*}}\\
    &=
    \sqrt{1-(\zeta^*)^{-1}}\int_0^{\pi/2}\frac{d\theta}{\sqrt{1-\frac{\cos^2\theta}{\zeta^*}}}\\
    &=
    \sqrt{1-(\zeta^*)^{-1}}K(\frac{1}{\sqrt{\zeta^*}}).
    \label{eq:example_ellip}
\end{align}

In obtaining the above, we have only used the definition of the elliptic integral $K(z)$. Arguing similarly for the rest of the terms in $I$ (save the last two elliptic integrals), $I(1-\zeta^*)$ becomes

\begin{multline}
    I(1-\zeta^*)
    =
    \left[\sqrt{1-(\zeta^*)^{-1}}K(\frac{1}{\sqrt{\zeta^*}}) - 2iK(\frac{1}{\sqrt{1-(\zeta^*)^{-1}}})\right]\\
    \times
    \left[
    \frac{1}{\sqrt{1-(\zeta^*)^{-1}}}E(\frac{1}{\sqrt{\zeta^*}})
    +2i\left(E(\frac{1}{\sqrt{1-(\zeta^*)^{-1}}})-K(\frac{1}{\sqrt{1-(\zeta^*)^{-1}}})\right)
    \right].
\end{multline}

To make further progress, we now use the reciprocal modulus relations for elliptic integrals, keeping in mind that $\zeta$ lies in the lower half-plane \cite{Fettis1970}. In particular, these relations state that given some complex $k'$ lying in the lower half-plane, we have $E(1/k') = [E(k')+iE(k)-k^2K(k')-ik'^2K(k)]/k'$ and $K(1/k') = k'[K(k')-iK(k)]$, where $k = \sqrt{1-k'^2}$. In our case, we take $k = 1/\sqrt{\zeta^*}$, or equivalently $k' = \sqrt{1-(\zeta^*)^{-1}}$. The above product then becomes
after some work

\begin{equation}
    I(1-\zeta^*)
    =
    \left[K(\frac{1}{\sqrt{\zeta^*}}) + 2iK(\sqrt{1-(\zeta^*)^{-1}})\right]
    \left[
    E(\frac{1}{\sqrt{\zeta^*}})-2i\left(E(\sqrt{1-(\zeta^*)^{-1}})-K(\sqrt{1-(\zeta^*)^{-1}})\right)
    \right].
\end{equation}

Using the same method we used to obtain Eq. (\ref{eq:example_ellip}), we have $K(\sqrt{1-(\zeta^*)^{-1}}) = \sqrt{\zeta^*}K(\sqrt{1-\zeta^*})$. Therefore, $I(1-\zeta^*)$ can be written as

\begin{equation}
    I(1-\zeta^*)
    =
    \left[K(\frac{1}{\sqrt{\zeta^*}}) + 2i\sqrt{\zeta^*}K(\sqrt{1-\zeta^*})\right]
    \left[
    E(\frac{1}{\sqrt{\zeta^*}})-2i\left(E(\sqrt{1-(\zeta^*)^{-1}})-K(\sqrt{1-(\zeta^*)^{-1}})\right)
    \right].
\end{equation}

Upon comparing this with $I(\zeta)$, we see that $I(1-\zeta^*) = (I(\zeta))^*$. The desired lemma immediately follows.

\subsection{Expanding $M_{++}(z_\bq,\bq)$ and $M_{+-}(z_\bq,\bq)$ at small $\bq$ in the presence of the long-range Coulomb interaction}
\label{app:expansions}
In this section, we obtain expressions for $M_{++}(z_\bq,\bq)$ and $M_{+-}(z_\bq,\bq)$ at small $\bq$ in the presence of the long-range Coulomb interaction. We begin with $M_{++}(z_\bq,\bq)$, which we simplify at small $\bq$ using the method of Ohashi and Takada \cite{Ohashi1998}. Noting that $\chi_{22}(0,0) = -2/g$, we rewrite $1/g + \chi_{22}(z,\bq)/2 = (\chi_{22}(z,\bq)-\chi_{22}(0,0))/2 \equiv \delta\chi_{22}(z,\bq)/2$. With this, $M_{++}(z,\bq)$ becomes

\begin{equation}
    M_{++}(z,\bq)
    = \frac{1}{2(V_c^{-1}(\bq)-\chi_{33}(z,\bq))}
    \left[V_c^{-1}(\bq)\delta\chi_{22}(z,\bq) - \left(\delta\chi_{22}(z,\bq)\chi_{33}(z,\bq)+\chi_{23}(z,\bq)^2\right)\right].
\end{equation}

We now define $K = \delta\chi_{22}(z,\bq)\chi_{33}(z,\bq)+\chi_{23}(z,\bq)^2$, which we would like to expand to lowest order in $\bq$. To do so, we expand $\delta\chi_{22}$, $\chi_{33}$, and $\chi_{23}$ to second order in $\bq$. Following Ref. \cite{Ohashi1998}, we define $\eta = k q \cos\theta/m$ and $Z = (E_++E_-)/((E_++E_-)^2-z^2)$. After some algebra, we then obtain

\begin{align}
    \delta\chi_{22}(z,\bq)
    &\approx
    -z^2\int \frac{d^2p}{(2\pi)^2}
    \frac{1}{2E^2}\left(1-\frac{\eta^2\gap^2}{2E^4}-\frac{\xi q^2}{4mE^2}\right)Z
    +\int \frac{d^2p}{(2\pi)^2}\left(\frac{3\eta^2\gap^2}{8E^5}+\frac{\xi q^2}{8mE^3}\right)\\
    \chi_{33}(z,\bq)
    &\approx
    -\int \frac{d^2p}{(2\pi)^2}\frac{2\gap^2}{E^2}\left(1+\frac{\eta^2\xi^2}{2E^4}-\frac{\xi q^2}{4mE^2}\right)Z\\
    \chi_{23}(z,\bq)
    &\approx
    -i\gap z\int \frac{d^2p}{(2\pi)^2}\frac{1}{E^2}\left(1-\frac{\eta^2(\gap^2-\xi^2)}{4E^4}-\frac{\xi q^2}{4mE^2}\right)Z.
\end{align}

With these expressions, we find that $K$ simplifies to

\begin{equation}
    K \approx -2 \gap^2 \int \frac{d^2p}{(2\pi)^2}\frac{Z}{E^2}\int \frac{d^2p'}{(2\pi)^2}\left(\frac{3\eta'^2\gap^2}{8E'^5}+\frac{\xi'q^2}{8mE'^3}\right).
\end{equation}

Note that we have not expanded $Z$ in powers of $\bq$, as it turns out to be unnecessary. Using this, we find that $M_{++}(z,\bq)$ to lowest-order in $\bq$ is given by

\begin{equation}
    M_{++}(z,\bq)
    =
    \frac{1}{4\gap^2}
    \bigg[-\frac{z^2}{2}V_c^{-1}(\bq)
    +2 \gap^2 \int \frac{d^2p'}{(2\pi)^2}\left(\frac{3\eta'^2\gap^2}{8E'^5}+\frac{\xi'q^2}{8mE'^3}\right)\bigg].
\end{equation}

Although we are mainly concerned with the Higgs mode, we note that in the high-density limit, we can solve for the dispersion of the phase mode by solving $M_{++} = 0$. In this limit, we find that the phase mode is given by $\om_p(\bq) = \sqrt{2N_0 V_c(\bq) v_F^2\bq^2\expval{\cos^2\theta}}$. In three dimensions, this yields $\om_p(\bq) = \sqrt{4\pi n e^2/m}$, while in two dimensions, this yields $\om_p(\bq) = \sqrt{2\pi n e^2 q/m}$. In other words, the phase-mode becomes the plasmon.

Since we are not concerned with the plasmon, we continue and keep only the lowest-order term in $M_{++}(z,\bq)$ \footnote{In three dimensions, $V_c(\bq) \sim 1/q^2$, both terms in $M_{++}(z,\bq)$ are of the same order, and $M_{++}(z,\bq) = O(q^2)$}, which in two dimensions is given by

\begin{equation}
    M_{++}(z,\bq) \approx -\frac{z^2}{8\gap^2}V_c^{-1}(\bq).
\end{equation}

We now move on to $M_{+-}(z,\bq)$, which we expand as we did $M_{++}(z,\bq)$. We write

\begin{equation}
    M_{+-}(z,\bq)
    = \frac{1}{2(V_c^{-1}(\bq)-\chi_{33}(z,\bq))}
    \left[V_c^{-1}(\bq)\chi_{12}(z,\bq) - \left(\chi_{12}(z,\bq)\chi_{33}(z,\bq)+\chi_{23}(z,\bq)\chi_{13}(z,\bq)\right)\right].
\end{equation}

To evaluate this, we first expand $\chi_{12}(z,\bq)\chi_{33}(z,\bq)+\chi_{23}(z,\bq)\chi_{13}(z,\bq)$. We find

\begin{align}
    \chi_{12}(z,\bq)
    &=
    i z \int \frac{d^2p}{(2\pi)^2}\left(\xi-\frac{\eta^2\gap^2\xi}{2E^4}+\frac{q^2(\gap^2-\xi^2)}{8mE^2}\right)\frac{Z}{E^2}\\
    \chi_{13}(z,\bq)
    &=
    -\gap\int \frac{d^2p}{(2\pi)^2}\left(2\xi+\frac{\eta^2\xi(\xi^2-\gap^2)}{2E^4}+\frac{q^2(\gap^2-\xi^2)}{4mE^2}\right)\frac{Z}{E^2}.
\end{align}

Using this along with our previous expressions for $\chi_{33}$ and $\chi_{23}$, we find after some algebra,

\begin{equation}
    \chi_{12}(z,\bq)\chi_{33}(z,\bq)+\chi_{23}(z,\bq)\chi_{13}(z,\bq)
    =
    \frac{i z \gap^2}{2}\int \frac{d^2p}{(2\pi)^2}\frac{d^2p'}{(2\pi)^2}
    \frac{ZZ'}{E^2E'^2}\xi'\left(\frac{\eta^2}{E^2}-\frac{\eta'^2}{E'^2}\right)
\end{equation}

This expression is nominally $O(q^2)$. However, we are interested in the Higgs mode, where $z_\bq = 2\gap + O(q^2)$. In this case, $Z \sim 1/(z_\bq^2-(E_++E_-)^2)$ is singular near $\bq=0$. In fact, using the same methods of section \ref{app:technical_details}, one can show that the above expression is $O(q)$ rather than $O(q^2)$. The other expression in the numerator of $M_{+-}(z_\bq,\bq)$, i.e. $V_c^{-1}(\bq) \chi_{12}(z_\bq,\bq)$, goes as $O(q)$, since $\chi_{12}(z_\bq,\bq)$ is a constant in the $q=0$ limit. Since the denominator is $O(1/q)$, (coming from the $\bq$-dependence of $\chi_{33}$), we have in total $M_{+-}(z_\bq,\bq) = O(q^2)$. This result holds in both two and three dimensions. In three dimensions, $V_c^{-1}(\bq) \sim q^2$, and can be neglected compared with the other $O(q)$ term in the numerator of $M_{+-}(z_\bq,\bq)$.

\bibliography{main}

\end{document}